\documentclass[11pt]{article}
\usepackage{jheppub}
\pdfoutput=1

\usepackage{dsfont}
\usepackage{amsmath,amssymb,amscd,amsfonts,mathtools}

\usepackage[toc,page]{appendix}
\usepackage{epsfig}
\usepackage{epstopdf}
\usepackage{latexsym}
\usepackage{graphicx}
\usepackage{subfig}
\usepackage{booktabs}
\usepackage{bbm}
\usepackage{color}
\usepackage{physics}
\usepackage{tensor}
\usepackage{tikz}
\usepackage{tcolorbox} 
\usetikzlibrary{matrix}
\usetikzlibrary{decorations.markings,calc,shapes,decorations.pathmorphing,patterns}
\usetikzlibrary{positioning}
\makeatletter
\def\@fpheader{\relax}
\makeatother

\definecolor{markgreen}{RGB}{230,243,230}

\subheader{\begin{flushright}

\end{flushright}}

\title{Entropy of Radiation with Dynamical Gravity}

\author{
Carlos Perez-Pardavila$^{a}$}
\affiliation{$^a$Weinberg Institute, Department of Physics, University of Texas, Austin, TX 78712, USA.}

\emailAdd{
cjp3247@utexas.edu}

\abstract{We compute the subregion entanglement entropy for a doubly holographic black string model. This system consists of a non-gravitating bath and a gravitating brane, where we incorporate dynamic gravity by adding a DGP term. This opens up a new parameter directly extending previous work and raises an important question about unitarity. In this note we analyse which theories in this big parameter space, will have unitary entropy evolution, in particular, we will distinguish which of those will follow a Page curve.}

\begin{document}	
\maketitle
\flushbottom

%%%%%%%%%%%%%%%%%
\section{Introduction}
%%%%%%%%%%%%%%%%%

Over the last few years, several new results have addressed and attempted to resolve a long-standing paradox about information and black holes. One important tool utilised in these recent developments is the Ryu-Takayanagi  prescription \cite{Ryu:2006bv,Ryu:2006ef,Hubeny:2007xt,Faulkner:2013ana,Lewkowycz:2013nqa,Engelhardt:2014gca}. In essence, it exploits the AdS/CFT correspondence \cite{Maldacena:1997re,Gubser:1998bc,Witten:1998qj} enabling to perform complicated entropy computations in the strongly conformal field theory by identifying them with area computations on the $AdS$ bulk. 

Boundary conformal field theories or BCFTs are $d$-dimensional strongly coupled conformal field theories on manifolds with boundary \cite{Cardy:2004hm}. The boundary, a $d-1$ manifold, is usually called the defect. Using holography once, we arrive at the so-called ``intermediate picture", in which the boundary is dual to $AdS_d$ gravity \cite{Karch:2000gx,Takayanagi:2011zk,Fujita:2011fp} localised in a Karch-Randall (KR) \cite{Karch:2000ct,Randall:1999vf} brane coupled to a non-gravitating bath where our CFT lives in \cite{Penington:2019npb,Almheiri:2019psf,Almheiri:2019hni,Almheiri:2019psy,Almheiri:2019yqk}. A second application of holography yields the full picture in which the brane is an end-of-the-world brane embedded in an $AdS_{d+1}$ space. The CFT now lives in the boundary of this bulk. These double descriptions are usually called ``doubly holographic” and throughout this note we will adhere to this convention.

In the past we have used this doubly holographic model to study the phase structure of the entropy in a subregion of a non-gravitating bath \cite{Geng:2021BHl}. This was itself an extension to the case where the bath was gravitating, forming a wedge-holographic model \cite{Geng:2020fxl}. In this note, however, we will keep the bath non-gravitating as in the previous study. In other words, we will consider a black string state in an $AdS_{d+1}$, in which we embed a KR brane. The advantage of this model is that it allows us to place a black hole in the background of the CFT, because the bulk theory is dual to a CFT thermal state in a non-gravitating eternal black hole background. On the other hand, the black string horizon is connected to the horizon of the black hole in the brane and the bath.

The main motivation behind this note is to continue analysing the time evolution of the entropy of a subregion of the bath in relation to the Page curve. The Page curve describes the unitary evolution one should expect entropy to follow according to our understanding of quantum mechanics \cite{lubkin1978entropy,Page:1993df}. In particular, we have previously observed the emergence of a Page curve from the bulk perspective by comparing two competing minimal area surfaces or \textit{RT surfaces} at a classical level. The early time entropy is generally controlled by a horizon penetrating surface, namely the \textit{HM surface} \cite{Hartman:2013qma} which under some circumstances is then replaced by a lower area surface called the \textit{island surface} whose area is constant \cite{Almheiri:2019yqk}. When this happens at some finite time, we say that the evolution of the entropy follows a Page curve. The latter type of surfaces anchor on the brane, enclosing the so-called \textit{islands} on it. The whole surface is the union of the island surface and its partner on the thermofield double side. These have been previously studied in \cite{Ling:2020laa,KumarBasak:2020ams, Emparan:2020znc,Caceres:2020jcn,Caceres:2021fuw,Deng:2020ent, Krishnan:2020fer,Balasubramanian:2020coy,Balasubramanian:2020xqf,Manu:2020tty,Karlsson:2021vlh,Wang:2021woy,Miao:2021ual,Bachas:2021fqo,May:2021zyu,Kawabata:2021hac,Bhattacharya:2021jrn,Anderson:2021vof,Miyata:2021ncm,Kim:2021gzd,Hollowood:2021nlo,Wang:2021mqq,Aalsma:2021bit,Ghosh:2021axl,Neuenfeld:2021wbl,Geng:2021iyq,Balasubramanian:2021wgd,Uhlemann:2021nhu,Neuenfeld:2021bsb,Kawabata:2021vyo,Chu:2021gdb,Kruthoff:2021vgv,Akal:2021foz,KumarBasak:2021rrx,Lu:2021gmv,Omiya:2021olc,Ahn:2021chg,Balasubramanian:2021xcm,Li:2021dmf,Kames-King:2021etp,Sun:2021dfl,Hollowood:2021wkw,Miyaji:2021lcq,Bhattacharya:2021dnd,Goswami:2021ksw,Chu:2021mvq,Arefeva:2021kfx,Shaghoulian:2021cef,Garcia-Garcia:2021squ,Buoninfante:2021ijy,Yu:2021cgi,Nam:2021bml,He:2021mst,Langhoff:2021uct,Ageev:2021ipd,Pedraza:2021cvx,Iizuka:2021tut, Miyata:2021qsm,Gaberdiel:2021kkp,Uhlemann:2021itz,Collier:2021ngi,Hollowood:2021lsw,emparan2021holographic,omidi2021entropy,bhattacharya2021bath,Merna}, and reviewed in \cite{Almheiri:2020cfm,Raju:2020smc,Raju:2021lwh,Liu:2020rrn,Nomura:2020ewg,Kibe:2021gtw}

It is clear that whenever these island surfaces exist and depending on their relative area to the HM surface, we can either have constant for all times or initially rising Page curves. However, when such surfaces don’t exist this story breaks and we lose any hope of describing evolution as unitary. This is for example the case of empty $AdS_{d+1}$ where the existence of such surfaces depends on a single parameter called the \textit{critical angle}.

This problem with unitarity is resolved if we include a black string in the bulk. As explained above, in this case we found  that an island surface always exists \cite{Geng:2021BHl}. This existence is independent of the two parameters in the theory, which are the anchoring point in the CFT, namely $\Gamma$ and the brane angle $\theta_b$. Therefore, one could classify the two types of entropy curves (i.e. constant-for-all-times or initially rising Page curve) in a phase diagram by studying the area difference between the island surface and the HM surface. This phase diagram is driven by the Page angle, which is defined as the angle for which the island surface and the HM surface have the same area. The critical angle still appears in this picture, since below the critical angle, island surfaces exist only above some anchoring point in the brane, called the critical anchor. Similar entropy phase structures have also been studied analytically in lower dimensions \cite{Geng_2022bhba,Geng:2021wcq}.

Following these results, one may wonder if they are still true in the case where one includes dynamical gravity on the brane. In lower dimensions this was achieved by including JT gravity on the brane \cite{Almheiri:2019hni}. In higher dimensions we can mimic this by including an Einstein-Hilbert term in the brane action similar to Dvali-Gabadadze-Porrati (DGP) gravity in an $AdS$ background; this was already done for empty $AdS$ \cite{Chen:2020uac}, for topological black holes \cite{Chen:2020hmv} and for the wedge black string model \cite{asRong-Xin_Miao1,Rong-Xin_Miao2,Liu_2022}. This opens a new parameter that our theory depends on, extending our previous result. This begs the question of whether the parameter space in this family of theories is limited. These limitations may arise in the same way as we observed in empty $AdS$, where the lack of RT-surfaces meant that entropy couldn't possibly evolve unitarily for some brane angles. Therefore, our first task will be to classify the structure of RT surfaces and compute their areas. In doing this we will not only address the unitarity problem, but also check when we have a Page curve. 

In particular, \cite{asRong-Xin_Miao1,Rong-Xin_Miao2} assert that for some part of parameter space, the wedge holographic model does reproduce the Page curve, while maintaining the massless graviton. Although we do not study this model here, we refer the reader to future work \cite{GengDGP}.

Before we summarise these results let us observe that in all our previous studies \cite{Geng:2021BHl,Geng_2022jt,Geng:2020fxl,Geng:2021hlu}, the natural choice of boundary conditions on the brane were Neumann boundary conditions. Without a DGP term these conditions translate to right-angle anchoring of the RT surface on the brane. An important consequence of the boundary conditions in the two-brane scenario is that the only surface which satisfied them on both branes was the horizon. This leads to the trivial result that the entropy between the defect and its thermofield double does not evolve with time (for empty $AdS$). Although we do not study the two-gravitating-brane picture in this note, we observe that relaxing the Neumann boundary conditions can enable new RT surfaces for this case which were not previously studied. In this former paper \cite{Geng:2020fxl}, as well as in the case of a non-gravitating bath, time-dependence came from those surfaces anchoring in the non-gravitating region, which is the case we are going to study in this note.

\paragraph{Overview of the Paper.} In section \ref{Section2} we present the setup of the system studied here. We remind the reader about double holography and the one-brane black string model. We then proceed to introduce dynamical gravity on the brane at the action level and derive the corresponding equations of motion to find the RT surfaces in section \ref{Section3}. In section \ref{Section4} we compute and present the corresponding areas and the phase structure. The note concludes in section \ref{Section5} with some remarks.

%%%%%%%%%%%%%%%%%
\section{Double Holography and Dynamical Gravity} \label{Section2}
%%%%%%%%%%%%%%%%%

\begin{figure}
    \centering
    \includegraphics[width=\linewidth]{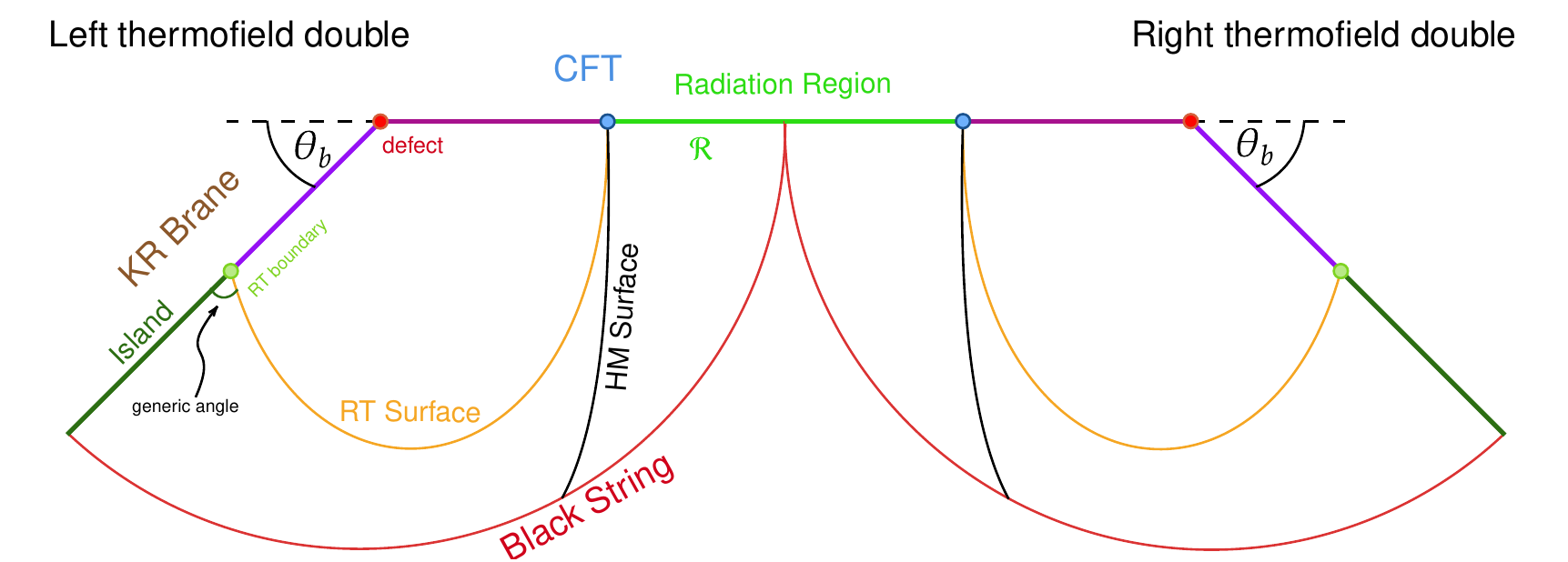}
    \caption{Cartoon of one copy of the KR braneworld considered in this note. The region behind the Karch-Randall brane has been excised. Two copies of this braneworld are glued along the brane. The intersection of the brane with the RT surface is not necessarily at right angles and will be determined by equation \eqref{eq:DGPBC}.}
    \label{fig:Setup}
\end{figure}

In this section we give a brief overview of the holographic model and the black string setup we will use in this paper, depicted in figure \ref{fig:Setup}. Since the setup is mostly the same as we have been using in the past few papers \cite{Geng:2020fxl,Geng:2021BHl}, we refer the unfamiliar reader to those papers. The new ingredients in this study will be covered in the latter parts of this section, so the familiar reader may want to skip to that part.

%%%%%%%%%%%%%%%%%
\subsection{Doubly holographic model and the black string}
%%%%%%%%%%%%%%%%%

The main setup we want to study in this note is the KR braneworld model. This model has a good description in terms of three equivalent pictures \cite{Karch:2000ct,Karch:2000gx,Almheiri:2019hni,Takayanagi:2011zk,Fujita:2011fp}:
\begin{itemize}
\item \textbf{Boundary:} $d$-dimensional CFT with a $(d-1)$-dimensional defect ($\text{BCFT}_d$ \cite{Cardy:2004hm,McAvity:1995zd}).
\item \textbf{Bulk:} Einstein gravity in an asymptotically $\text{AdS}_{d+1}$ spacetime containing an $\text{AdS}_d$ Karch-Randall (KR) brane \cite{Karch:2000ct}. This KR brane also has dynamical gravity.
\item \textbf{Intermediate:} $d$-dimensional CFT coupled to ``dynamical" gravity on the $\text{AdS}_d$ brane \cite{Karch:2000ct}, with transparent boundary conditions between the brane at infinity and a nongravitating $\text{CFT}_d$ (the \textit{bath}) on half space.
\end{itemize}

One of the most important consequences of the transparent boundary conditions between the brane and the non-gravitating bath is that the graviton on the brane acquires a mass \cite{Porrati:2001gx,Porrati:2002dt,Porrati:2003sa}. This can be seen from the boundary description, since these boundary conditions lead to the non-conservation of the stress tensor. This in turn picks up an anomaly dimension, which translates to the mass of the graviton \cite{Aharony:2006hz} \footnote{It is a standard fact in $AdS/CFT$ that the anomalous dimension of operators in the boundary appears as a mass term in the $AdS$ equations.}. The mass of the graviton can be related to the brane angle (i.e. it is roughly quadratic in the brane angle) for small angles \cite{Miemiec:2000eq,Schwartz:2000ip} and to the $AdS_{d+1}$ length scale for large angles. This means that most computations of the Page curve in $d>2$ at a semiclassical level have been performed in models with a massive graviton  \cite{Geng:2020qvw}. Since we are using the same model, this note and all computations and results presented here have the same feature. We remark that in \cite{asRong-Xin_Miao1,Rong-Xin_Miao2} the authors claim they're able to reproduce the Page curve evolution in massless gravity.

Now that we have seen the general idea of coupling a gravitating region to a non-gravitating bath and the three equivalent descriptions that holography provides, it is time to include the black string. From the bulk perspective, we start with a $(d+1)$-dimensional black string environment in which we embed a $d$-dimensional KR brane. Setting the $AdS_{d+1}$ radius to $1$ we can describe the bulk geometry with the metric,
\begin{equation}
\begin{split}
ds^2 &= \frac{1}{u^2 \sin^2\mu}\left[-h(u) dt^2 + \frac{du^2}{h(u)} + d\vec{x}^2 + u^2 d\mu^2\right],\\
h(u) &= 1 - \frac{u^{d-1}}{u_h^{d-1}},
\end{split}\label{eq:BS}
\end{equation}
\noindent where $t \in \mathbb{R}$, $u > 0$, $0 < \mu < \pi$, and $\vec{x} \in \mathbb{R}^{d-2}$. These coordinates slice the $AdS_{d+1}$ space into $AdS_d$ subspaces, one for each constant $\mu$. The black string induces a black hole in each slice, one of which is the KR brane, i.e. the $\mu = \theta_b$ slice. In particular, by the $AdS/CFT$ correspondence, this setup is a gravitating black-hole coupled to a non-gravitating black-hole background in the bath. 

Typically, the black string features a thermodynamic instability, called the Gregory-Laflamme instability. In global $AdS$ this instability is controlled by a combination of two of the parameters in the theory, the $AdS$ radius and the horizon distance $u_h$. Considering large enough black strings (compared to the horizon distance) we avoid this instability. For the case at hand, by considering the Poincar\'e patch we automatically avoid this, i.e. we only consider large black holes. Small black holes will be in considered in \cite{Merna}.

We are interested in finding the entanglement entropy between a subregion $\mathcal{R}$ (the radiation region) in the bath and its complement $\bar{\mathcal{R}}$ at a fixed time slice (which we choose to be $t=0$). We will use the usual RT prescription in which we first find all extremal surfaces $\Sigma$ satisfying the homology constraint
\begin{equation}
\partial\Sigma = \partial \mathcal{R} \cup \partial \mathcal{I}.
\end{equation}
The second boundary term in the homology constraint, written in terms of the island $\mathcal{I}$, allows for surfaces to cross the brane into the other KR braneworld.

%%%%%%%%%%%%%%%%%
\subsection{Area functional}
%%%%%%%%%%%%%%%%%

Now we wish to explicitly find the surfaces satisfying the constraint above. As we will see shortly, and at the level of finding the RTs, the DGP term will affect the boundary condition so we will start by not writing it explicitly. The final step will be to extremise the set of these surfaces and pick the minimal area one. 

Looking at the metric, \eqref{eq:BS}, we can exploit the translational symmetry along $x_i$ to parameterise the potential RT surfaces as $u(\mu)$. Substituting this into our metric, for a constant $t=0$ slice, we can write an area density functional
\begin{equation}
\mathcal{A} = \int_{\theta_b}^{\pi} \frac{d\mu}{(u\sin\mu)^{d-1}} \sqrt{u^2 + \frac{u'(\mu)^2}{h(u)}}, \label{eq:AF}
\end{equation}
We observe that $u_h$ can be written as an overall factor of $u_h^{2-d}$ in the integral above, meaning that it scales out of the problem. Thus, we can without loss of generality set $u_h=1$. Furthermore, we remark that we will be, mostly, omitting the word ``density" throughout this note

Our objective now is to extremise this functional and if the set of such extremising surfaces contains more than one element, to pick the minimal area one. This is done by solving the Euler-Lagrange equation (i.e. the equation of motion) and imposing appropriate boundary conditions. This equation can be written explicitly as
\begin{equation}
\begin{split}
u'' = -(d-2) u\,h(u) + (d-1) u' \cot\mu \left(1 - \frac{\tan\mu}{2} \frac{u'}{u\,h(u)} + \frac{u'^2}{u^2\,h(u)}\right)
- \left(\frac{d-5}{2}\right) \frac{u'^2}{u}.
\end{split} \label{eq:eomODE}
\end{equation}
Until now, this is apparently the same as we have done in the previous study, so let's see what changes now.

%%%%%%%%%%%%%%%%%
\subsection{DGP term}
%%%%%%%%%%%%%%%%%

The area of the RT surface to compute entanglement entropy arises from evaluating the Ricci scalar on a solution with a conical defect, required by the replica trick \cite{Lewkowycz:2013nqa}. Therefore, following \cite{Chen:2020uac}, by introducing a Ricci scalar (via an Einstein-Hilbert term) on the brane, namely induced by the bulk gravity, we can get an additional contribution to the entanglement entropy, called the DGP term. The full action on the brane will thus be,
\begin{equation}
    S = -T \int d^d x \sqrt{-\tilde g} + \frac{1}{16 \pi G_b} \int d^d x \sqrt{-\tilde g} \tilde R
\end{equation}
where the $\tilde g$ is the induced metric on the brane, $\tilde R$ is the corresponding Ricci scalar, $T$ is the brane's tension \footnote{This is related to the brane angle via $T =\frac{(d-1)}{4 \pi G} [\cos \theta_b + \frac{\lambda_b}{2} (d-2) \sin^2 \theta_b]$. The author would like to thank Hao Geng for highlighting this important point.} and $G_b$ is the Newton's constant on the brane which parameterises the strength of gravity there (i.e. the DGP term). With this, the entanglement entropy computed from the area of the RT surface is
\begin{equation}
    S_{EE} = \frac{1}{2} \left( \frac{2}{4G} \mathcal{A} + \frac{1}{4G_b}  \mathcal{A}_b \right) = \frac{1}{8G} (2\mathcal{A} + \lambda_b \mathcal{A}_b) \label{eq:SEE}
\end{equation}
where $\mathcal{A}$ is calculated in \eqref{eq:AF} and $\lambda_b = \frac{G}{G_b}$ \footnote{We will occasionally omit the subindex and write $\lambda$ instead of $\lambda_b$ to simplify notation.} \footnote{G is the bulk's Newton's constant.}. There is also a factor of $2$, which comes from the fact that we have two copies of spacetime in our KR braneworld. The factor of $\frac{1}{2}$ comes from the $\mathbb{Z}_2$ orbifolding, although this is a conventional choice. The latter is the independent parameter that we are going to be free to tune and it is the main new ingredient in this note. The term $\mathcal{A}_b$ is the area density of the point where the RT surface intersects the brane, it is thus given by
\begin{equation}
    \mathcal{A}_b = \left. \frac{1}{(u \sin \mu)^{d-2}} \right|_{u=u_b,\mu = \theta_b} \label{eq:AF2}
\end{equation}
where $u_b$ is the $u$-coordinate where the brane intersects the RT surface. This is the last piece that we need in order to derive the correct boundary conditions for the problem at hand, so let's proceed to this now.

The equation of motion \eqref{eq:eomODE} will not be affected by the boundary term $\mathcal{A}_b$. In this way, abbreviating the equation of motion by $EOM$ and extremising
\begin{equation}
\begin{split}
    0 \equiv & 2 \delta \mathcal{A} + \lambda_b \delta \mathcal{A}_b=- \int_{\theta_b}^\pi d\mu (\delta u) (EOM) \\
    & +\left .\frac{2 \delta u}{(u \sin \mu)^{d-1}}\frac{u'}{h(u) \sqrt{u^2 + \frac{(u')^2}{h(u)}}}\right |_{\theta_b}^{\pi} + \left. \frac{(2-d)\lambda_b \delta u}{u^{(d-1)} (\sin \mu)^{d-2}} \right|_{u=u_b,\mu = \theta_b}
\end{split}
\end{equation}
For Neumann boundary conditions on the brane we must set the second line to zero, evaluated at the boundary on the brane, namely
\begin{equation}
    u' = \left. \pm (2-d) \frac{u \, \lambda_b \, h(u) \, \sin \mu}{\sqrt{4-(d-2)^2 h(u) \, \sin^2\mu \, \lambda^2}} \right|_{u=u_b,\mu = \theta_b} \label{eq:DGPBC}
\end{equation}
If we further assume $d>2$, and since $0<\mu< \pi/2$ then the sign of $u'$ is fixed to be the same as the sign of $-\lambda_b$, by requiring that it indeed solves the Neumann boundary conditions. We observe that the square root in the denominator of \eqref{eq:DGPBC} means that $\lambda_b$ can only take values in $(-\frac{2}{(d-2)\sin \theta_b \sqrt{h(u_b)}},\frac{2}{(d-2)\sin \theta_b \sqrt{h(u_b)}})$, for which $u'$ can take any value, for fixed $\theta_b$ and $u_b$. 

For the rest of this note we will fix $d=4$ \footnote{This means, when comparing with \cite{Chen:2020uac}, that the induced Newton's constant $G_{RS} = (d-2) G/2$ is the same as the bulk Newton's constant $G$ and hence their $\lambda_b$ is the same as our $\lambda_b$. This is not true in general dimensions (or general radius curvature).}. Following the discussion in \cite{Chen:2020uac} we can also write the induced Newton's constant on the brane as
\begin{equation}
    \frac{1}{G_{eff}} = \frac{1}{G} (1+\lambda_b)
\end{equation}
This means that at the special value of $\lambda_b = -1$ gravity on the brane becomes infinitely strong and we should be careful when analysing the results. Furthermore, although we will analyse all values of $\lambda_b$, if $\lambda_b \leq -1$, the effective Newton's constant $G_{eff}$ is negative and so the theory is non-physical. 

As we can see from the equation of motion \label{eomODE}, solving it analytically is quite difficult so we are going to follow the same approach as in \cite{Geng:2020fxl} and solve it numerically, using a shooting method.

We will now proceed to present some results and therefore (in addition to fixing $d=4$) we fix $u_h=1$.  The latter will not affect any result, since this is an overall scale in the functional which doesn't affect the equations of motion.

%%%%%%%%%%%%%%%%%
\subsubsection{The $\lambda$-parameter}
%%%%%%%%%%%%%%%%%

Geometrically we can think of the boundary condition given by $u’$, as the angle at which the RT surface intersects the brane. For instance, when $u’(\theta_b)=0$ we recover our old right-angle condition, for example if $\lambda_b=0$. However for other values of $\lambda_b$ the situation is more complicated since, by looking at \eqref{eq:DGPBC}, we also have to consider $u_b$ and $h(u_b)$.

For example, if we fix the shooting point to $u_b=0$, then $u'=0$ independently of the value of $\lambda_b$. We observe that this will not always be possible because the critical anchor $u_{crit}$ also plays a role in this story. We remind the reader that the \textit{critical anchor} \cite{Geng:2020fxl} is the lowest point on the brane from which it is not possible to reach the bath, by shooting from it. In other words, RT candidates are defined in the region of the brane $(u_{crit}, u_h)$, which receives the name of \textit{atoll}. For $\lambda_b = 0$ the critical anchor is zero for a brane angle $\theta_b \geq \theta_c$, where $\theta_c$ is the critical angle. For general $\lambda_b$ the critical anchor is non-zero even above the critical angle and hence in general we are not always able to shoot from $u_b=0$.

On the other hand if we fix $u_b = u_h$ then equation \eqref{eq:DGPBC} implies that $u'(\theta_b)=0$, for any value of $\lambda_b$. This means that if we shoot to the bath from the horizon, we will stay at the horizon and we will always reach the bath. If the surface penetrates the horizon and does not reach the bath on the TFD side from which it was launched, its area will represent increasing entropy over time and will never result in a Page curve. Therefore, for the parameter combinations where this occurs, the entropy evolution in our theory will not be unitary and we will reject this theory. In other words, in addition to the reality condition we imposed a priori on $\lambda_b$, we will further impose that the corresponding RT surfaces reach the bath. 

\begin{figure}
    \centering
    \includegraphics[width=0.9\linewidth]{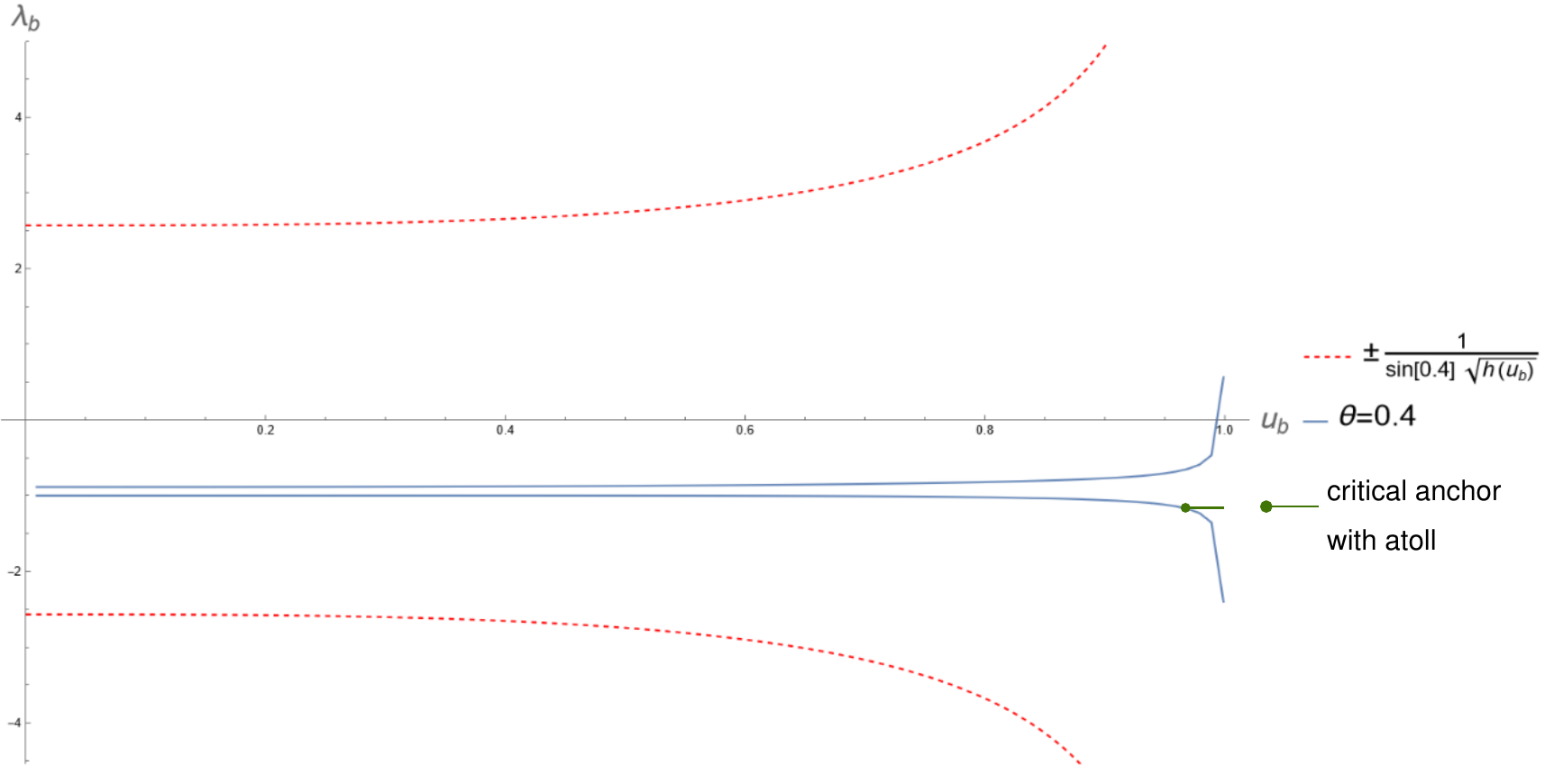}
    \caption{Bound on the parameters $\lambda_b$ and $u_b$ for which we can reach the bath. The region inside the red dashed curves shows the theoretical bound we derived above. The region inside the blue solid curves is the numerical region we obtained from solving the differential equation. For one $\lambda_b$ a critical anchor is marked by a green dot and the shooting points making up the atoll are marked with a green line.}
    \label{fig:Lvsu01}
\end{figure}

Doing this numerically gives us a critical anchor and an atoll for each angle and $\lambda_b$. Surprisingly the bounds imposed by this are lower than the theoretical reality bounds, as seen in figure \ref{fig:Lvsu01}. 

\begin{figure}
    \centering    \includegraphics[width=\linewidth]{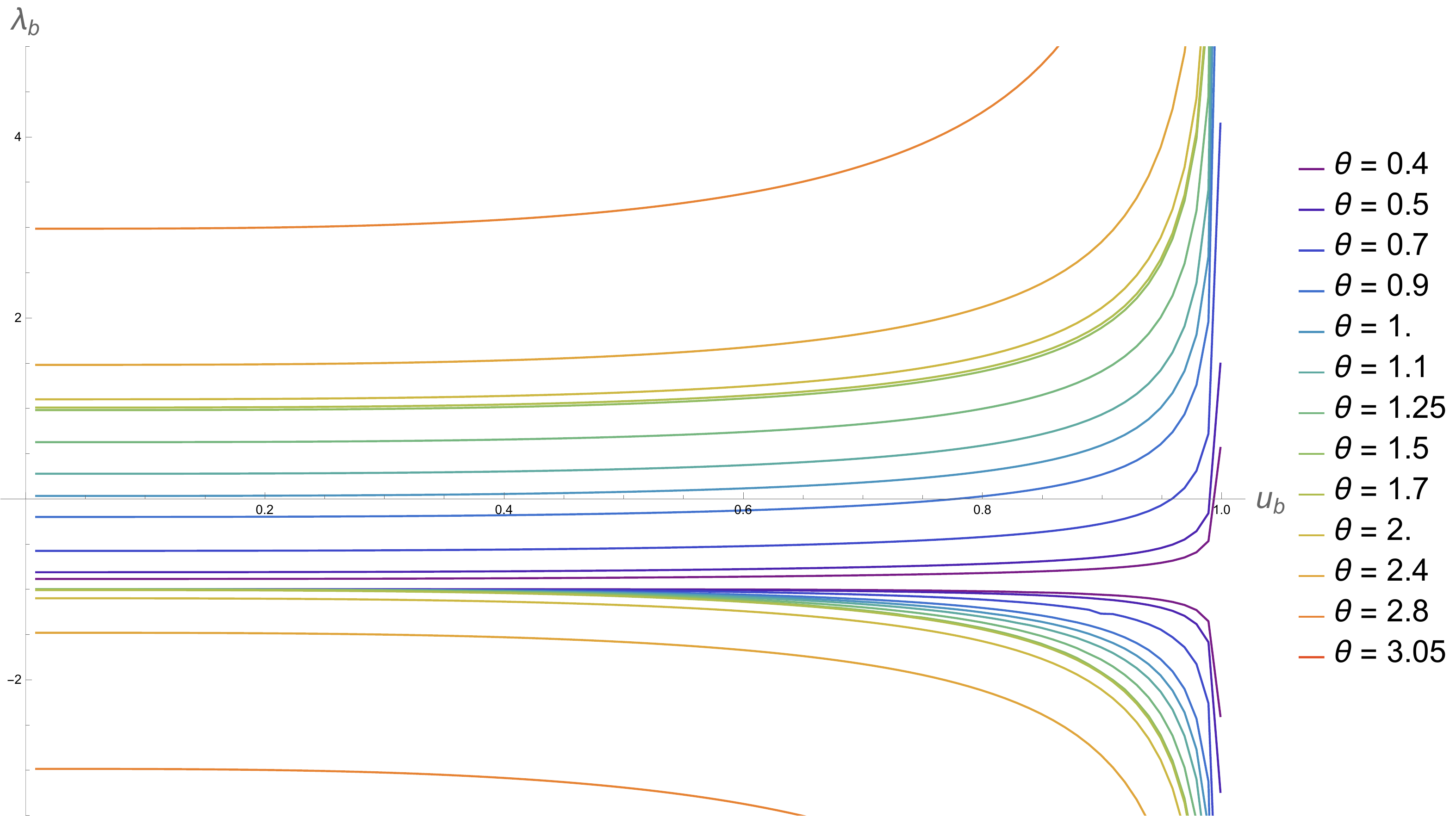}
    \caption{Bound on the parameters $\lambda_b$ and $u_b$ for which we can reach the bath for several brane angles $\theta_b$. The region inside curves of the same color represent the values of $\lambda_b$ and $u_b$ for which we can reach the bath. This region gets wider as we increase $\theta_b$.}
    \label{fig:Lvsu02}
\end{figure}

This figure shows how for fixed values of $\lambda_b$ we get a critical anchor and a corresponding atoll which changes as we change $\lambda_b$. This generalises our old story of critical anchors, which is now both $\lambda_b$ and $\theta_b$ dependent (see figure \ref{fig:Lvsu02}). Furthermore, we observe that there is a range of $\lambda_b$ values, for which shooting anywhere on the brane, we can reach the bath. In other words, there is a range where the critical anchor shrinks to zero. As an example, for the case shown in figure \ref{fig:Lvsu01}, this is $\lambda_b \in (-0.884,-0.999)$

By performing the computation for multiple values of the brane angle, we can see how the range of values gets wider as the $\theta_b$ increases, shown in figure \ref{fig:Lvsu02}. We also notice that as the brane angle increases, the lower bound curves decrease, while the upper bound curves increase. For small values of the anchoring point u, when $\abs{\lambda_{b}}<1$ this increase (or decrease) is not substantial and so all curves accumulate near $\lambda_b= \pm 1$. These are somewhat special values, at $\lambda_b = -1$ the effective Newton's constant on the brane becomes infinite. On the other hand, at $\lambda_b=1$, the value of the brane and the bulk Newton's constants match numerically, $G_b =G$ \footnote{We remark that these do not have the same dimensions and hence this equality is just a numerical one.}. In particular, for the angles considered here, we will see that $\lambda_b=-1$ is also special when we study the RT structure. For $\theta_b = \theta_{c} \approx 0.98687$ the lower bound curve of the range crosses entirely the $\lambda_b=0$ axis. This is consistent with previous results in \cite{Geng:2020fxl} since for angles above the critical angle the atoll should cover the whole brane. Additionally, for angles below the critical angle, the critical anchor monotonically decreases as the angle increases for any fixed $\lambda_b \geq -0.884$ and in particular for $\lambda_b=0$.

%%%%%%%%%%%%%%%%%
\section{RT Structure} \label{Section3}
%%%%%%%%%%%%%%%%%

In this section we are going to explore the different types of Ryu-Takayanagi (RT) surfaces we can have for a fixed set of parameters, that is for a fixed anchoring point in the bath $\Gamma$, a fixed value for $\lambda_b$ and a fixed brane angle $\theta_b$. We begin by describing the Hartman-Maldacena (HM) surface, which penetrates the Einstein-Rosen bridge and extends into the thermofield-double side. The other surfaces are going to be different versions of island surfaces, which anchor on the brane and cross into the second copy of spacetime in our KR braneworld.

\subsection{HM surface}

To study the Hartman-Maldacena surface we are going to change the parametrisation of our surface, such that $\mu = \mu(u)$. In this way the area functional becomes:
\begin{equation}
\mathcal{A} = \int \frac{du}{(u \sin\mu)^3} \sqrt{\frac{1}{h(u)}+u^2 \mu'(u)^2} \ .\label{eq:AFmu}
\end{equation}
where, since we know the surface doesn’t intersect the brane, the DGP term doesn’t affect the solutions to the Euler-Lagrange equations of this functional. To determine the boundary conditions at the horizon we can, in general, consider the Hartman-Maldacena surface anchoring at $\Gamma$ on one side of the thermofield double and $\tilde \Gamma$ at the other side. For simplicity and following \cite{Geng:2021BHl}, we will choose the symmetric case $\Gamma = \tilde \Gamma$. This imposes Dirichlet boundary conditions on the bath as with any other RT surface.

The other boundary is at the horizon, so we need to determine what the conditions are here. In order to do so we are going to assume this surface is smooth across the Einstein-Rosen bridge, in other words, and since we assumed the HM anchors at the same distance from the defect on both sides of the TFD, that it has no ``kinks". In adapted (tortoise-like) coordinates, $dr = \frac{du}{u\sqrt{h}}$, this can be mathematically written as: $\mu'(r=0)=0$

Following \cite{Geng:2021BHl}, we can either expand the solution to the equation of motion in this coordinates, or translate to our old parametrisation $\mu(u)$. If we do the latter, we note that the expansion will contain half integer powers of $(u_h-u)$ \footnote{This is because $r \sim \sqrt{u_{h}-u}$}. After doing this, we uniquely fix the solution by choosing the $\Gamma$ and the angle at which the HM surface crosses the horizon $\theta_{HM}$. This is equivalent (using our power expansion) as choosing the boundary conditions:
\begin{equation}
\mu(u_h \Gamma)=\pi, \hspace{1cm}
\mu(u_{h}) = \theta_{HM}, \hspace{1cm} \mu'(u_{h}) = \frac{2 \cot \theta_{HM}}{u_{h}},\label{eq:bdryHM}
\end{equation}

\subsection{Island surfaces}

In this section we are going to analyse the RT surfaces giving rise to islands, i.e. surfaces which anchor on the bath and the brane. In the doubly holographic models, islands are generally considered to be the regions on the brane going into the horizon, depicted in figure \ref{fig:Setup}. In previous work we found that, when there was no DGP term, the existence of RT surfaces was driven by a specific angle value, which we called the critical angle $\theta_c$. For angles $\theta<\theta_c$, there is a specific value of anchoring point, the critical anchor, on the brane below which we can never reach the bath \cite{Geng:2020fxl}. 

\begin{figure}
    \centering
    \includegraphics[width=0.9\linewidth]{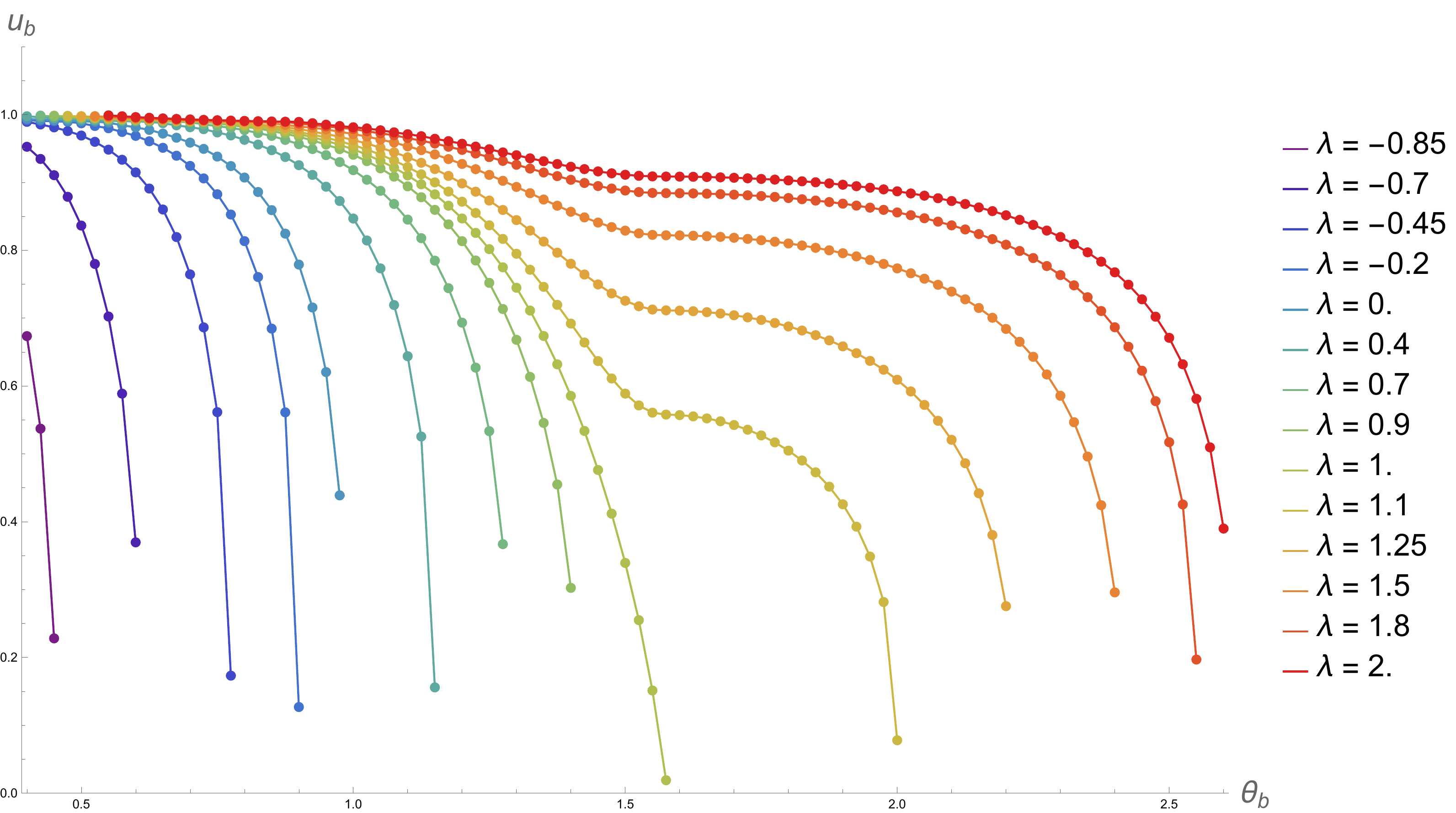}
    \caption{Critical anchors as a function of brane angle $\theta_b$ for several values of $\lambda_b > -1$.}
    \label{fig:CritAnch}
\end{figure}

\begin{figure}
    \centering
    \includegraphics[width=0.9\linewidth]{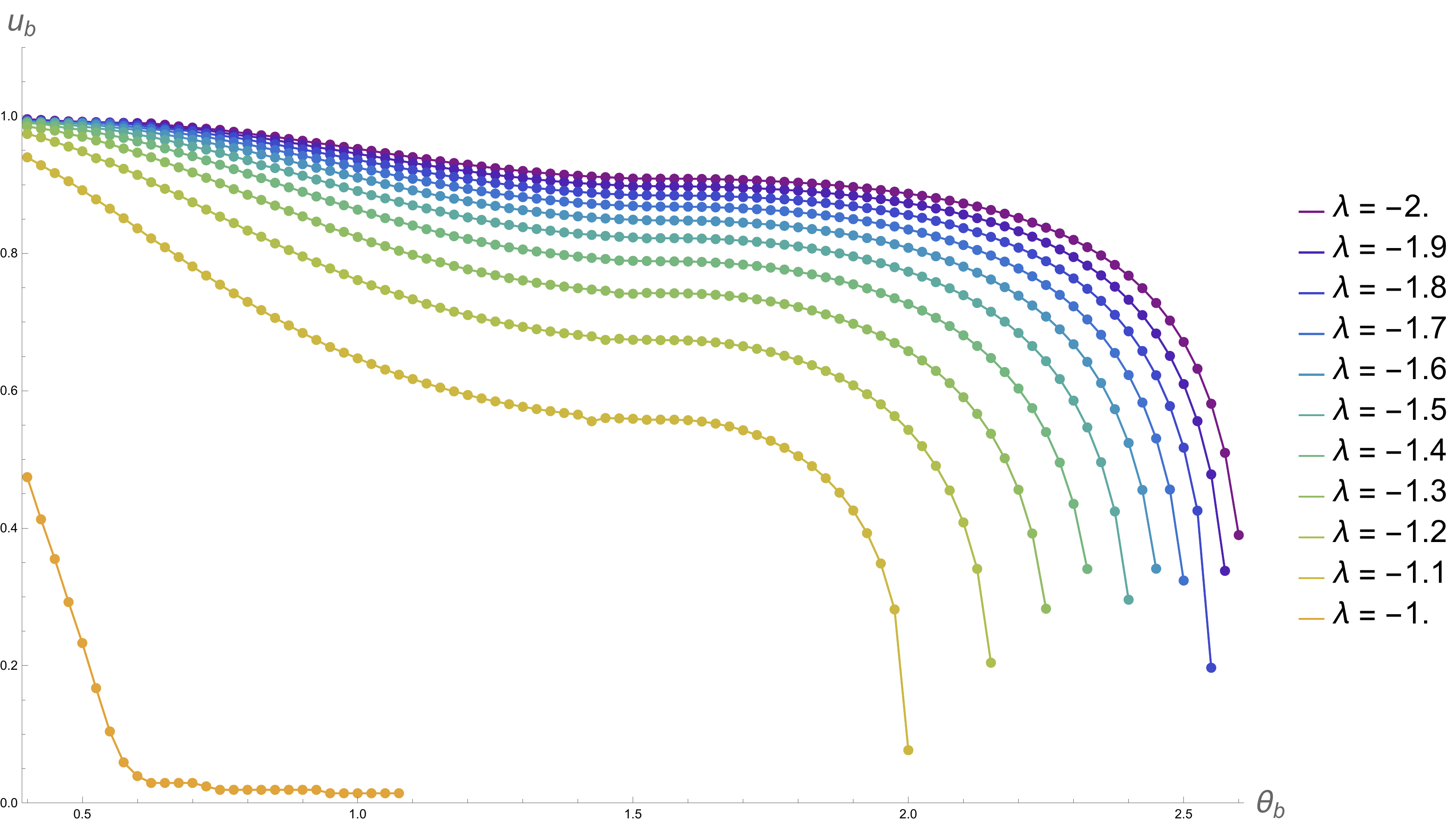}
    \caption{Critical anchors as a function of brane angle $\theta_b$ for several values of $\lambda_b \leq -1$.}
    \label{fig:CritAnch2}
\end{figure}

Something similar will occur in our setup; for a fixed angle and a fixed DGP coupling $\lambda_b$, there is a finite range of anchoring points from which we can reach the bath. In other words, we generalise the critical anchor slightly, where now this depends on these two parameters. We already noted this in figure \ref{fig:Lvsu02}, where we gave the minimum range of values for which we can reach the bath by shooting from anywhere in the brane. As we increase or decrease $\lambda_b$, we can reach the whole bath only by shooting from a section of the brane, namely the atoll. We can extract the critical anchor information and plot them as a function of $\theta_b$ for each $\lambda_b$, like we show in figure \ref{fig:CritAnch} for $\lambda_b>-1$ and figure \ref{fig:CritAnch2} for $\lambda_b<-1$. 

As we can see in figure \ref{fig:CritAnch} the critical anchors are monotonically decreasing with increasing brane angle or decreasing $\lambda_b$. There are two main behaviours we observe in this figure. When $1<\lambda_b<2$ the curves don't decrease with decreasing slope, but show a small bump. The $\lambda_b=1$ value is somewhat significant because it is where the Newton's constants on the brane and the bulk numerically match.

When $-1<\lambda_b<1$, we see that the critical anchors, all follow similar curves to our previous results, i.e. $\lambda_b=0$ \cite{Geng:2020fxl}. In fact, if we think of the critical angle as the angle at which the critical anchor shrinks to the defect $u_b=0$, then by following the curves in figure \ref{fig:CritAnch} to their intersection with the $\theta_b$-axis, we can also define a generalised critical angle. As we can see this decreases with increasing $\lambda_b$. Of course, one may think of the critical angle as determined in empty $AdS$ and hence there is no possible generalisation in this case. In the next section we are going to take this latter interpretation.

When $\lambda_b \leq -1$, the behaviour of the critical anchors in relation to $\lambda_b$ is reversed. As seen in figure \ref{fig:CritAnch2}, decreasing $\lambda_b$ increases the critical anchor, while the behaviour in relation to the brane angle remains unchanged. In this case, we also notice that the critical anchors seem to have a similar behaviour as when $1<\lambda_b<2$ with a slope that appears to be small (near zero) for some angles. This produces generalised critical angles above $\pi/2$ for $\abs{\lambda_b}>1$. When $\lambda_b=-1$, i.e. when $G_{eff}$ becomes infinite making the theory on the brane non-physical, the critical anchors approach $u_b=0$ asymptotically.

Another way of putting the above observations is that the size of the atoll increases when we increase the brane angle or when the value of $\lambda_b$ approaches $-1$, from above in figure \ref{fig:CritAnch} or from below in figure \ref{fig:CritAnch2}. If we view the degrees of freedom on the brane as redundant with those on the bath, in the limit when the effective Newton's constant is infinite, i.e. $G_{eff} \rightarrow \infty$ information on the brane gets delocalised. On the other hand when this ratio has a large absolute value information on the brane is localised near the horizon, i.e. the resulting size of the island is small and is located near the horizon.

\begin{figure}
    \centering
    \includegraphics[width=1
\linewidth]{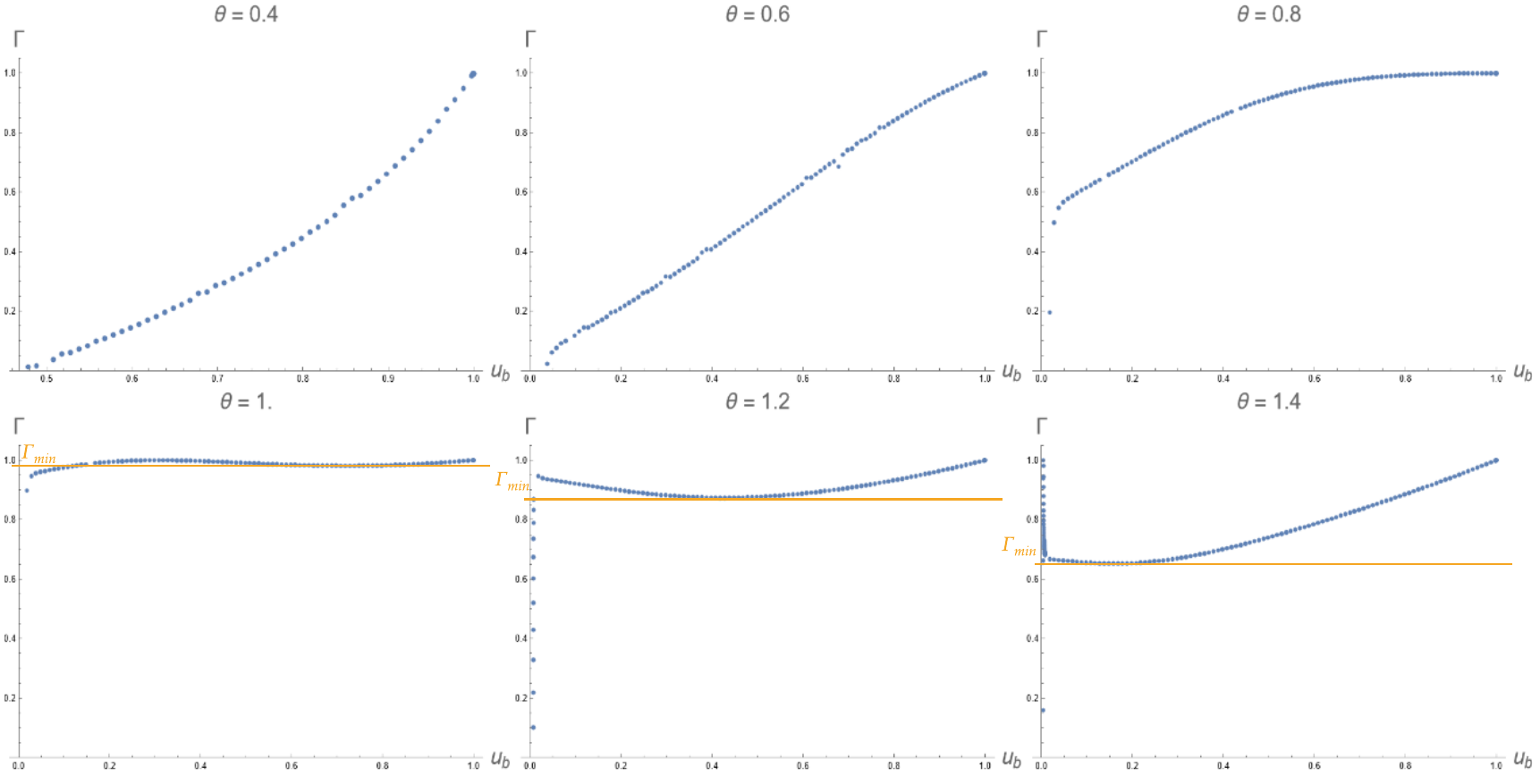}
    \caption{Plots of the shooting points on the brane $u_b$ vs the anchoring points on the bath $\Gamma$ for several values of $\theta$ for $\lambda=-1$. The yellow lines show the $\Gamma_{min}$ values.}
    \label{fig:AnchvsGamma}
\end{figure} 

\begin{figure}
    \centering
    \includegraphics[width=.7\linewidth]{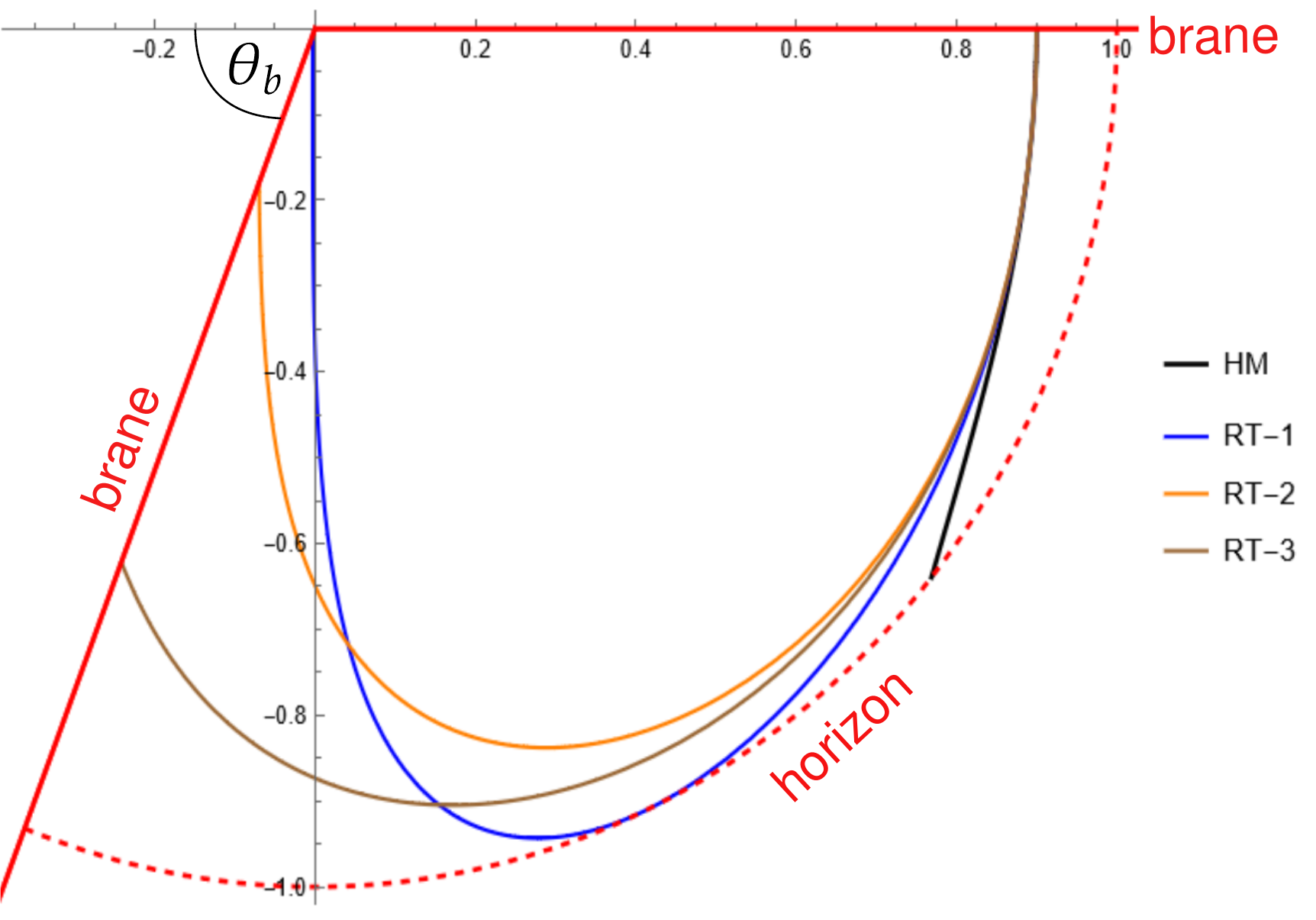}
    \caption{Polar plot of the three RT-candidates described above for $\Gamma=0.9$. The value brane angle is $\theta_b=1.2$, for which $\Gamma_{min}=0.873845$.}
    \label{fig:PolarPlot}
\end{figure}

Furthermore we can study in more detail how the anchoring point on the brane affects the anchoring point on the bath. When doing so we discover an interesting behaviour at $\lambda_b \leq -1$, which we depict in figure \ref{fig:AnchvsGamma}. We distinguish two behaviours. When $\theta_b \leq 0.825$ there is a one-to-one correspondence between shooting points on the brane, above some critical anchor, and an anchoring point $\Gamma$ on the bath. This is similar to the situation we observed for other values of $\lambda_b$. However when we increase the angle past this value, the critical anchor shrinks to the defect and there is a region close to the horizon, for which a single point in the bath may have up to three corresponding RT-candidates. This region is characterised by some $\Gamma_{min}$ above which these three surfaces exist and below which we have a unique RT candidate. This unique RT candidate is the one which has a smallest shooting point $u_b$ and which we call \textit{RT-1}. The next RT-candidate, ordering them by shooting point is what we call \textit{RT-2}. Finally the RT-candidate which is shot farthest from the defect receives the name of \textit{RT-3}. This situation is depicted in figure \ref{fig:PolarPlot}.

Moreover, the value of $\Gamma_{min}$ monotonically decreases with increasing brane angle $\theta_b$\footnote{see figure \ref{fig:Supp1} in appendix \ref{app:A}.}. Associated to this value there is a shooting point $u_{min}$, such that the point $(u_{min}, \Gamma_{min})$ is in the curves shown in figure \ref{fig:AnchvsGamma}. These $u_{min}$ points also decrease with increasing $\theta_b$. The important thing for the case at hand, is that following the RT prescription, when there is a competition between RT-candidates we need to choose the minimal area one. Thus, if there is some transition in the RT surface between the RT-candidates listed, this occurs for decreasing values of $\Gamma$. This is in fact what will happen as we will see shortly, once we compute the areas of the RT-candidates.

When $\lambda_b=-1.1$, the same three competing surfaces and the same behaviour is found for angles $\theta_b \geq 1.225$. When $\lambda_b = - 1.2$ this behaviour is found for angles $\theta_b \geq 1.4$ and when $\lambda_b = - 1.3$ it is found for $\theta_b \geq 1.525$. This poses a technical difficulty when computing areas for $\lambda_b \leq -1$ and angles above $\pi/2$. Furthermore, as we will see in the next section, the areas generally decrease with $\lambda_b$ for any given brane angle. Therefore, with the aim of characterising when the areas vanish and producing a phase diagram, we will decide to compute the areas up to a brane angle of $\pi/2$.

\section{Numerical Results} \label{Section4}

Now that we have given a detailed account of what the possible RT-candidates are we can proceed to compute their areas. As we mentioned above we will have two competing RT surfaces: the island surface and the HM surface. The former has a constant area in time while the latter has an area which grows in time, because of the growing Einstein-Rosen bridge. In this section we compute the area differences between the island surfaces and the Hartman-Maldacena surface 
\begin{equation}
    \Delta A (t) = A_{IS} – A_{HM} (t)  \label{eq:DeltaA}
\end{equation}
which is automatically finite in the UV. We perform the computation at $t=0$, since we know what the time-evolution of each term. We use the results from our previous section about the existence of RT surfaces to find the areas \footnote{Formally we are computing area differences given by \eqref{eq:DeltaA}, but in the text we will loosely refer to ``areas".}. Finally, from this data we obtain the entropy phase structure for the present case.

The situation where $\lambda \leq -1$ brings a new challenge, since for certain angles we have several competing surfaces, so we will also need to be careful here to compute the smaller area surface. 

\subsection{Area difference for fixed parameters}

Before proceeding we make a small philosophical distinction about the parameters. Although there is no a priori hierarchy between the three parameters and we are free to choose, $\lambda_b$, $\theta_b$ and $\Gamma$, we would like to think about the first two as in some sense fixed before $\Gamma$. What we mean is that the brane angle and the gravity on the brane are fixed in the particular universe we study. However, the location on the bath where we measure the entropy depends on the location of the observer. In other words, an observer in a particular universe can always move further from the defect to choose another region on the bath to measure the entropy. That observer cannot change the brane angle nor the strength of the gravity on the brane unless that observer changes universes. Although we'll vary all of the parameters in our computations, we are going to do so by fixing $\lambda_b$ and $\theta_b$ and plotting the areas as a function of $\Gamma$.

Methodologically, we have decided to compute the areas for angles $\theta \in (0.4,\pi/2)$ (in steps of $\theta = 0.025$) \footnote{We remark that the tensionless branes are generally not at $\theta_b=\pi/2$, given that the tension vanishes at some angle depending on $\lambda_b$, which can be obtained from the relation between brane tension and angle quoted above. The choice of this range of angles is based on practical reasons, namely that the region of interest for this note lies in this range, as we will see form the phase diagram.}. The parameter $\lambda_b$ will run in the range $(-2,2)$, in steps of $\lambda_b = 0.05$. It's worth noting that this range includes values that correspond to theories that are considered non-physical. However, for the sake of completeness, we will still consider these values in our analysis. As we noted above, the particular value of $\lambda_b$ may limit the range of shooting points on the brane, i.e. the size of the atoll, but in terms of the position on the bath $\Gamma$, we will still cover the whole bath, so $\Gamma \in (0,1)$.

\subsubsection{$\lambda=-1$}

We now present some of the technical difficulties, in particular for the $\lambda=-1$ case, where the situation is a bit more intricate. We have noticed above that for a fixed $\Gamma$ we can have up to three competing RT surfaces. We have ordered these according to the shooting point on the brane: RT-1, the one closer to the defect, RT-2, the one in the middle and RT-3, the one closer to the horizon (see figure \ref{fig:PolarPlot}). In figure \ref{fig:AnchvsGamma} we can see that when the angle is $\theta \leq 0.825$, there is a unique surface for each shooting point on the brane and hence there is no competition. Furthermore, when $\Gamma<\Gamma_{min}$, there is also a unique surface, so again, this is the smallest area surface. However, when we increase the angle above $0.85$ two surfaces, RT-2 and RT-3 appear for $\Gamma>\Gamma_{min}$. Let us analyse each of the areas independently to explain this situation.

For RT-1, the area is positive near $\Gamma=0$ and diverges as $\Gamma \rightarrow 1$. The shooting point for RT-1 is located on the brane, close to the $AdS$ boundary, and it is therefore not regulated. This might lead one to expect a positive divergence for this area, but this is compensated by subtracting a similar DGP contribution, in \eqref{eq:SEE}. When $\Gamma=1$ the area blows up again, because here the HM has large negative renormalised area. This behaviour was noted in \cite{Ryu:2006ef}. This competition between the non-boundary area and the DGP term, produces a nearly constant area for every angle, for this surface.

For RT-2 and RT-3, we see a similar behaviour, reaching a smallest area at the point where we shoot directly to $\Gamma_{min}$, since this shooting point is closer to the defect. More precisely, RT-2 and RT-3 have monotonically increasing area for $\Gamma>\Gamma_{min}$. However, in this range of $\Gamma$s, RT-3 has slightly smaller area, only equating RT-2 precisely at $\Gamma_{min}$, where these surface are degenerate \footnote{See figure \ref{fig:Supp2} in appendix \ref{app:A} which illustrates this explanation.}. This rules out RT-2 as a competing surface. Nevertheless, these surfaces have decreasing area with increasing angle and hence, immediately after appearing they will dominate over RT-1. Both surfaces have the same divergence when $\Gamma \rightarrow 1$, which we observed in RT-1 and is what we would expect for any of these.

Collecting these results, we have the following behaviour in the area for a fixed angle as a function of $\Gamma$. Below $\theta_b=0.85$, the area (determined by the area of RT-1) is slightly positive close to the defect and then increases with $\Gamma$. It diverges again for $\Gamma$ close to the horizon because of the Hartman-Maldacena surface. When the angle is at or above $0.85$ this behaviour is maintained at $\Gamma<\Gamma_{min}$, being RT-1 the dominant surface. At $\Gamma_{min}$, there is a transition in which RT-3 becomes dominant, with a smaller area and keeps dominating all the way up to the stage where $\Gamma$ reaches the horizon and the area diverges. This produces a jump from a larger area RT-1, to a smaller area, RT-3 which looks discontinuous. Although here we won't resolve this discontinuity, we suspect that sampling more points in the bath can show a steep but smooth transition between these surfaces. This would produce two points for which $\Delta \mathcal{A} =0$. However the conservative approach we'll take here is to think of this jump as discontinuous. This is similar to the discontinuity observed in \cite{Geng:2020fxl,Merna}, when the tiny islands take over. As we can see in figure \ref{fig:AnchvsGamma}, increasing the angle means that $\Gamma_{min}$ is decreased and appears for lower shooting points $u_b$. In other words RT-2 and RT-3 appear ``sooner" and thus this jump also occurs sooner.

\subsubsection{Area difference as a function of $\lambda_b$}

\begin{figure}
    \centering
    \includegraphics[width=1\linewidth]{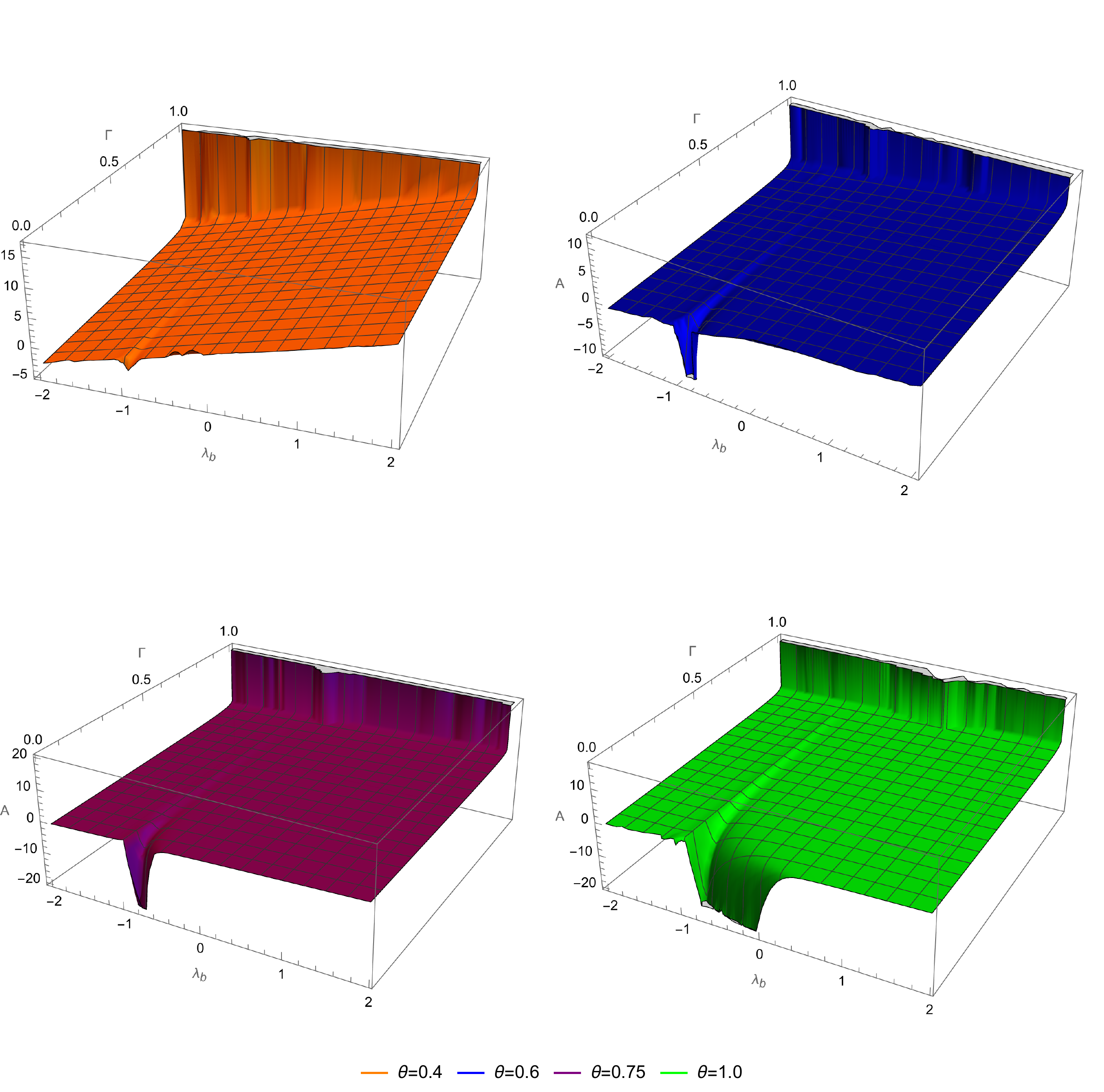}
    \caption{Area of the RT surface as a function of $\Gamma$ and $\lambda_b$ for angles between $\theta=0.4$ and $\theta=1$.}
    \label{fig:AvsGvsL1}
\end{figure}

\begin{figure}
    \centering
    \includegraphics[width=1\linewidth]{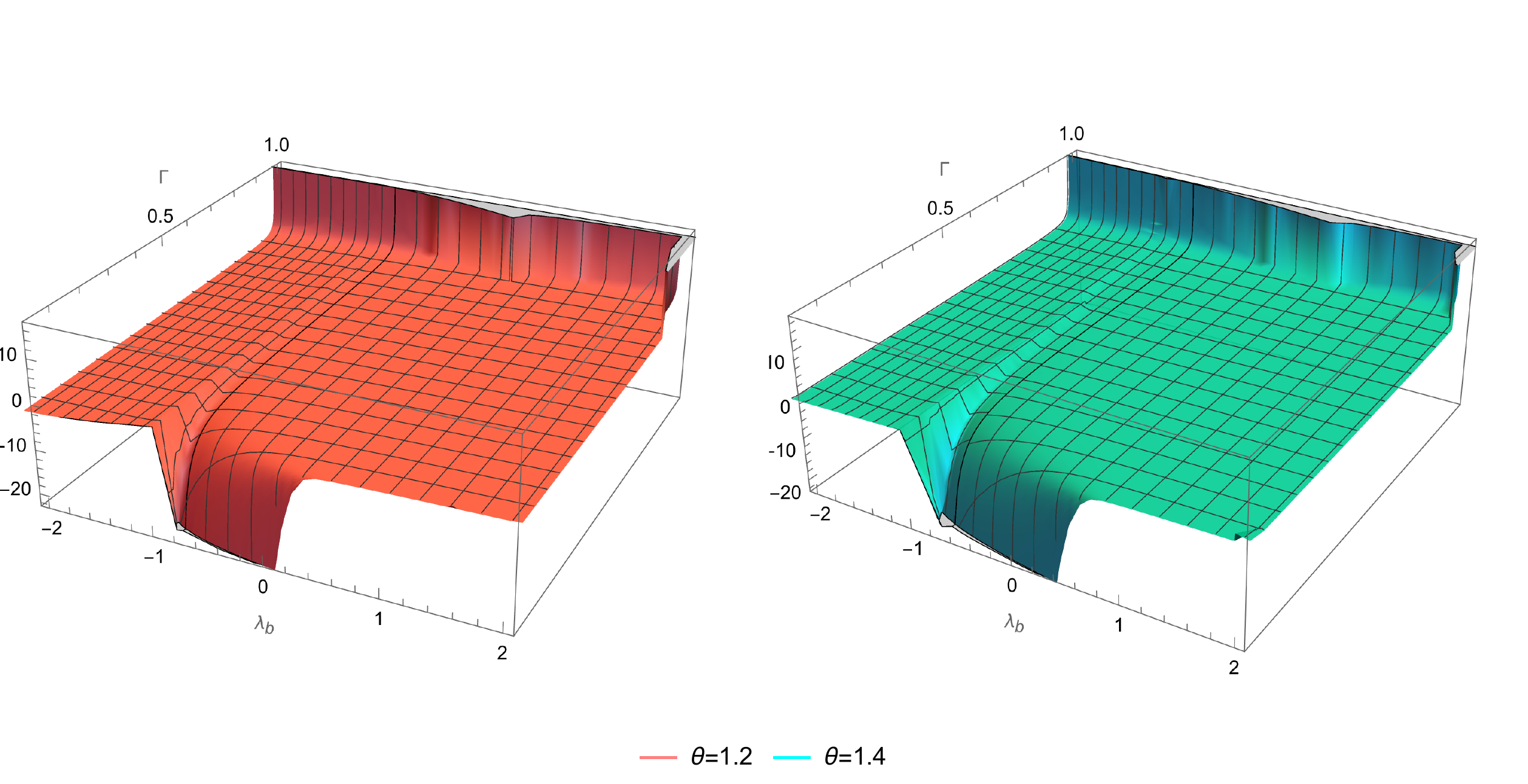}
    \caption{Area of the RT surface as a function of $\Gamma$ and $\lambda_b$ for angles $\theta=1.2$ and $\theta=1.4$.}
    \label{fig:AvsGvsL2}
\end{figure}

Here we look closer at the numerical results, depicting only a representative subset of the areas computed. These show the main features we observed in our computations. In particular, we can plot the area difference $\Delta \mathcal{A}$ for $t=0$ as a function of $\Gamma$ for several $\lambda_b$-values. This produces a 2-D surface for each angle. Below, in figures \ref{fig:AvsGvsL1} and \ref{fig:AvsGvsL2} we just show a few of these surfaces.

As we can see in figures \ref{fig:AvsGvsL1} and \ref{fig:AvsGvsL2}, the area increases monotonically with $\lambda_b$ until we reach $\lambda_b=-1$. Here we observe that the areas near the defect, $\Gamma=0$, are positive, as we described above and which will be more evident when we present these areas for fixed $\lambda_b$s. In figure \ref{fig:AvsGvsL1}, for $\lambda_b$ between $-1$ and $0$ we see how the near defect area becomes increasingly negative as we increase the angle. This behaviour is exacerbated until we reach $\theta_b=1$, for which the negative areas also appear closer to $\lambda_b=0$. In figure \ref{fig:AvsGvsL2}, this trend is maintained even for positive $\lambda$s. To explain this we recall that for $\lambda_b=0$ we recover our results in \cite{Geng:2020fxl}. Here we observed this divergence, which was caused by infinitesimal surfaces which wrapped the defect, namely the tiny islands. For non-zero values of $\lambda_b$, we can now have a non-zero critical anchor, even above the critical angle. Hence we don't obtain these tiny islands as limits of any surface, namely we cannot continuously shoot to the defect. However, the critical anchor gets smaller as $\lambda_b \rightarrow -1$ from below (see figure \ref{fig:CritAnch2}) and so, when we have large negative $\lambda_b$ ($\lambda_b \leq -1$) the RT-surface has a finite large shooting point $u_b$ and doesn't diverge to negative infinity. As we increase $\lambda_b$ the critical anchor sinks into the defect and the tiny islands take over. For small negative values of $\lambda_b$, we are subtracting a large negative area from an already negative term (tiny islands) by the DGP contribution as given in equation \eqref{eq:AF2}. Since the range of $\lambda_b$s for which the critical anchor shrinks into the defect increases with angle, this produces an increasingly pronounced dip in figures \ref{fig:AvsGvsL1} and \ref{fig:AvsGvsL2} near $\Gamma=0$. Something similar occurs for larger values of $\lambda_b$, where again the appearance of a critical anchor truncates the existence of tiny islands, preventing the area difference to diverge to negative infinity.

This behaviour suggests that any definition of the critical angle that depends on the existence of tiny islands, or as mentioned before, on the critical anchor sinking to the defect, will also be affected by the inclusion of the DGP term. However, our focus is on comparing our parameters to the critical angle of empty $AdS$ without a DGP term, and thus we will not define the critical angle in this manner.

Furthermore, away from the $\Gamma=0$ defect the areas are monotonically increasing with one exception. Although it is less noticeable in the other angles, in figure \ref{fig:AvsGvsL2}, for $\theta_b=1.4$ and $\lambda_b=-1$, the area monotonically increases until it reaches the $\Gamma_{min}$ we described before. At this point we have a discontinuity corresponding to the transition between RT-1 and RT-3. Away from the $\lambda_b=-1$ value the area is finite and monotonically increases with $\Gamma$ for every angle. In all cases, the area diverges to infinity as $\Gamma \rightarrow 1$. Above we mentioned that this divergence was caused by the fact that the HM surface has a renormalised area which diverges to $-\infty$. In other words, the explanation of this divergence comes from the fact that in \eqref{eq:DeltaA} we are subtracting a $-\infty$ term. Another way to see this divergence is using the fact that when $\Gamma$ gets close to the horizon, the corresponding HM shrinks to a point. This means that it ``cuts off" less of the RT surface which still picks up a divergence.

We recall that that a negative area difference $\Delta \mathcal{A}$, means that the island surface is smaller than the HM surface, which generally grows with time. Therefore, we can summarise these observation by: an observer sitting close to the defect for $\lambda_b \geq -1$, will (depending on the angle) measure a constant entropy curve. Furthermore, because of the competing surfaces at larger angles, this is also true for an observer sitting closer to the horizon. We remind the reader that exactly at $\lambda_b=-1$, the effective Newton's constant on the brane, $G_{eff}$, becomes infinite and hence the theory is considered non-physical. As we increase $\lambda_b$ the area differences will become positive and the observer will be able to measure a non-constant entropy curve.

\subsubsection{Area difference as a function of $\theta_b$}

\begin{figure}
    \centering
    \includegraphics[width=1\linewidth]{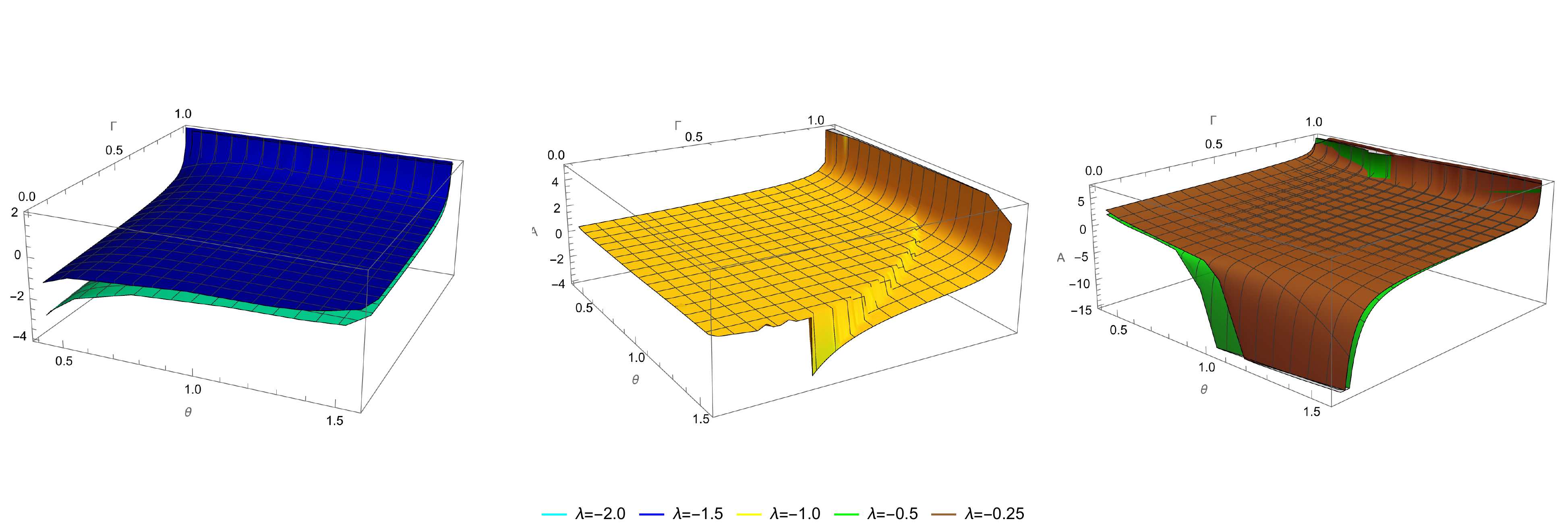}
    \caption{Area of the RT surface as a function of $\Gamma$ and $\theta_b$ for $\lambda_b < 0$.}
    \label{fig:AvsvsGvsTh1}
\end{figure}

\begin{figure}
    \centering
    \includegraphics[width=1\linewidth]{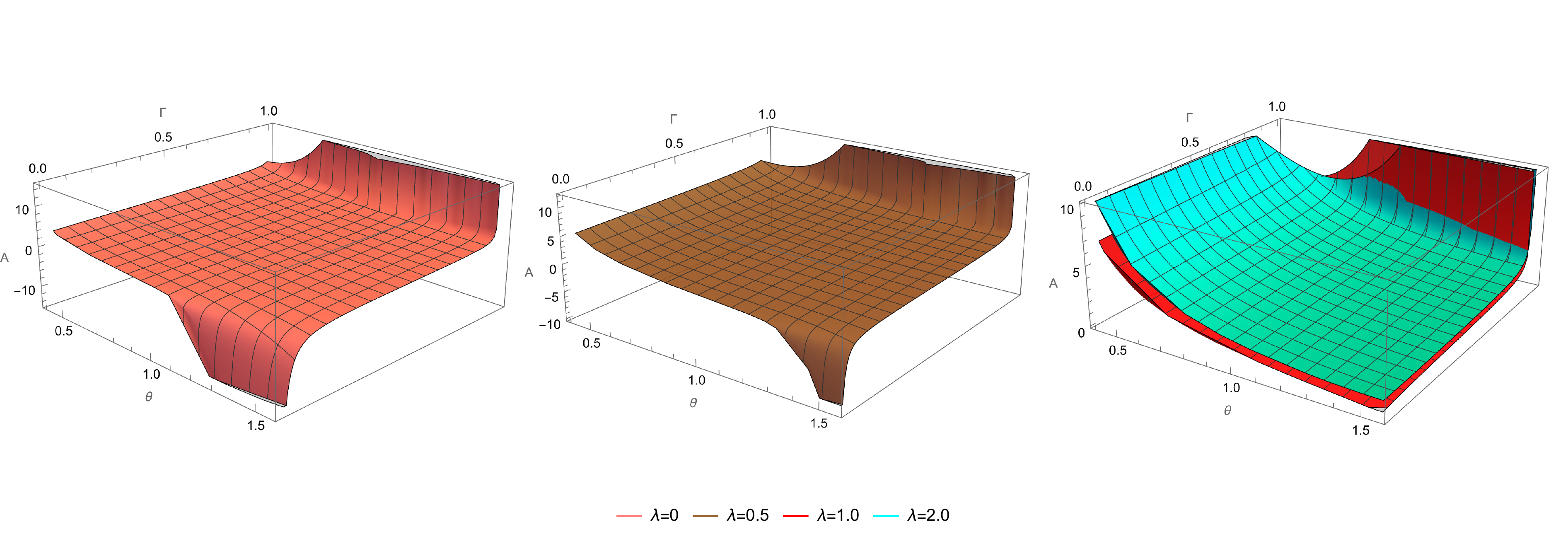}
    \caption{Area of the RT surface as a function of $\Gamma$ and $\theta_b$ for $\lambda_b \geq 0$.}
    \label{fig:AvsvsGvsTh2}
\end{figure}

To get some more intuition and to show some of the things we described above, we can similarly fix the $\lambda_b$ value and plot $\Delta \mathcal{A}$ vs $\Gamma$ for several angles. In this way each 2D surface corresponds to one $\lambda_b$-value. We can see these surfaces in figures \ref{fig:AvsvsGvsTh1} and \ref{fig:AvsvsGvsTh2}

In figure \ref{fig:AvsvsGvsTh1} we show some of the surfaces obtained for $\lambda_b \leq 0$. When $\lambda_b$ is sufficiently large and negative (panel on the left of the figure) we see that the area is finite independently of the angle \footnote{We expect the areas to diverge as the brane angle approaches $\theta_b \rightarrow 0$, from it becoming the $AdS$ boundary, but this is outside the range of values computed in this note}. This is what we observed above and it is because there is a non-zero critical anchor which, even when $\Gamma=0$, keeps our surface anchored away from the defect. The area monotonically increases with the brane angle $\theta_b$ and $\lambda_b$. As before, the area is divergent when $\Gamma \rightarrow 1$. 

In the centre figure, the areas are approximately constant with brane angle, but increasing with $\Gamma$ for most of parameter space. However for larger angles we observe the transition between RT-1 and RT-3. The latter surface has monotonically increasing area with $\Gamma$, which is also appreciated in this panel. Furthermore, we observe how this transition occurs at decreasing $\Gamma_{min}$ as we increase the brane angle. We remind the reader that thus far, these theories have $G_{eff}<0$ and hence can be considered as non-physical.

On the right panel of figure \ref{fig:AvsvsGvsTh1}, the areas still increase with $\Gamma$, but the trend with respect to $\theta_b$ is reversed, namely now we have decreasing areas with brane angle. This behaviour with respect to the two parameters will be maintained for all $\lambda_b >-1$ in figure \ref{fig:AvsvsGvsTh2}. The other new ingredient is the tiny island effect which appears when $\Gamma \rightarrow 0$ and $\theta>\theta_{crit}$. The dependence of the critical anchor on $\lambda_b$ can clearly be seen by the fact that the areas diverge faster to $-\infty$ for angles larger than $\theta_c$ as $\lambda_b$ approaches zero from below. The leftmost panel of figure \ref{fig:AvsvsGvsTh2} shows the case where the DGP coupling is turned off, $\lambda_b=0$ and this matches the results in \cite{Geng:2020fxl}. 

Comparing these figures we see how the tiny islands are somewhat tamed by the DGP coupling. However as we can see by comparing the first and third panel of \ref{fig:AvsvsGvsTh1} the DGP term has a greater effect on the smaller angles, reversing the monotonic behaviour of the area as a function of brane angle. This is evident from equation \eqref{eq:AF2} which is proportional to $1/\sin{\theta_b}^2$ and since the anchors are roughly fixed in a small range, as we would expect for small angles. We will comment on the signs of the areas below.

In general, the areas for any $\lambda_b$ and $\theta_b$ can be positive or negative so we can ask the question about when the areas are precisely zero, this is what we proceed to do now.

\subsection{Phase structure}

We have presented a detailed account for the area differences as a function of the different parameters in our theory; $\Gamma$, $\theta_b$ and $\lambda_b$. Now that we have observed the different trends we can analyse the entropy phase structure given in this parameter space. We recall that the sign of $\Delta \mathcal{A}$ will determine what surface dominates at $t=0$. Since the HM has a linear area evolution with time, a positive $\Delta \mathcal{A}$ means that the entropy follows a Page curve evolution. However, when $\Delta \mathcal{A}$ is negative the constant area island surface is the RT-surface whose area is proportional to the entropy and hence the entropy is just constant. Therefore, the question of when $\Delta \mathcal{A}$ vanishes is a question about when we have an entropy evolution following a Page curve. We remark that both situations satisfy unitarity, since we already discarded the non-unitary theories, that is how we limited the parameter space in the first place.

Here we present the main result in this note which is the entropy phase structures diagrams. These are curves showing the values of $\Gamma$ and $\theta_b$ for which $\Delta \mathcal{A} = 0$ for fixed $\lambda_b$ values.

\begin{figure}
    \centering
    \includegraphics[width=1\linewidth]{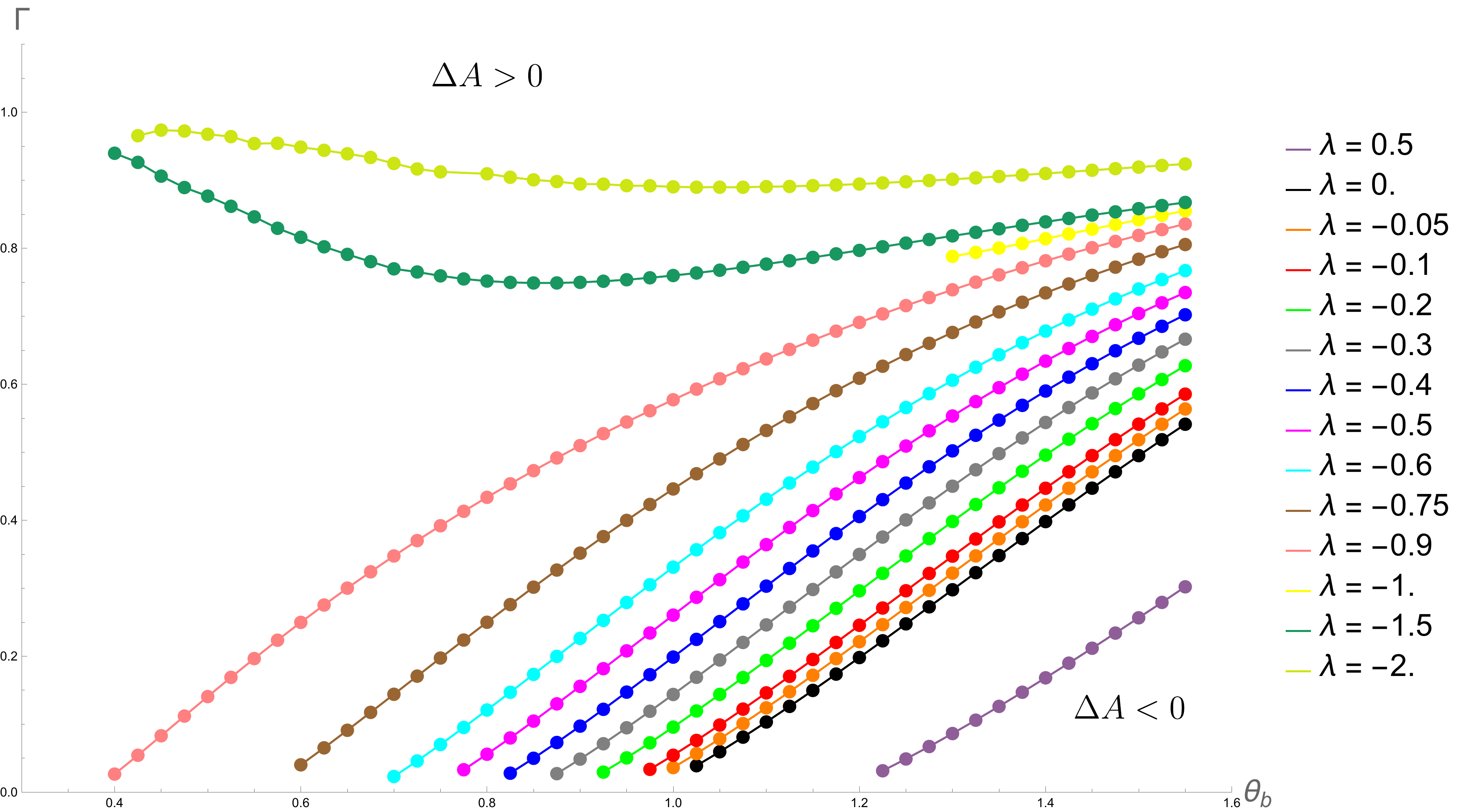}
    \caption{$\Gamma$ and $\theta_b$ for which $\Delta \mathcal{A}=0$ for several values of $\lambda_b$. The horizon is at $\Gamma=1$. For the first few curves the area is positive for a small range near the horizon, independently of the angle. Our previous results are reproduced by the black curve $\lambda_b=0$.}
    \label{fig:PhaseStructure}
\end{figure} 

From the plot in figure \ref{fig:PhaseStructure} we observe several different behaviours. When $\lambda_b$ is between $-2$ and $-1$ we observe that the curves in the phase diagram never intersect the $\theta_b$-axis. This means that the area never vanishes at the defect (for this range of angles), or in other words there is no proper definition of the Page angle, the angle for which the vanishing area surface has $\Gamma=0$. In fact, for these few first curves, the region where $\Delta \mathcal{A}>0$ is a smaller region of parameter space which is also close to the horizon. Since $\Delta \mathcal{A}<0$ means that there is a flat entropy curve (because the island surface starts dominant at $t=0$), this means that for most of the parameter space there is no Page curve. In particular when $\lambda_b = -2$ or $\lambda_b = -1.5$, we see the tendency of the curve to drop, but it is still mostly close to the horizon. Again we remark that these theories (when $\lambda_b \leq -1$) are unphysical from the start. However in the limit where $G_{eff} \rightarrow \infty$, or in other words when $\lambda_b \rightarrow 1$, we see that the competing surfaces described above only allows for vanishing areas in a small region of angles. Therefore, the corresponding $\lambda_b = -1$ curve (in yellow) is shorter than the other curves. In this case we can imagine a vertical line from the left end of this curve delimiting the $\Delta \mathcal{A} >0$ region from the $\Delta \mathcal{A}<0$ region.

When $\lambda_b>-1$, the curves intersect the $\theta_b$-axis and roughly half of parameter space will give us a page curve, $\Delta \mathcal{A}>0$ and half will have a constant entropy evolution. As we increase $\lambda_b$ the constant entropy region gets smaller. In fact when $\lambda_b>1$, we didn't observe negative areas and hence there is no vanishing area curves, although we suspect this will not be the case if we look at larger angles. This can also be seen in figure \ref{fig:AvsvsGvsTh2}. This means that the corresponding curves (which are not shown in the figure), would just be flat lines sitting on the $\theta_b$-axis.

We can also reformulate these observations in terms of what, in previous works, we called the constant entropy belt. This is the region of the bath containing anchoring points $\Gamma$, for which the area is $\Delta \mathcal{A}<0$. In the past, \cite{Geng:2021BHl}, this belt had a size which was monotonic with angle and this is what we observe here for $\lambda_b > -1$. In particular we observe this for what would reproduce our previous results $\lambda_b=0$. In this way we remark that this figure, directly extends figure 10 in \cite{Geng:2021BHl} However, when $\lambda_b=-2$ or $\lambda_b=-1.5$, this monotonicity is broken. We see the belt first decreases and then increases in both cases.

Finally, since some of the curves appear to intersect the $\theta_b$-axis, we can extract, by extrapolating these curves, a Page angle for each value of $\lambda_b$. We now proceed to do this extrapolation. This extrapolation will carry some intrinsic error, but to show the robustness of our method, we observe that for $\lambda_b=0$ the Page angle we obtain is $0.972$. This is only marginally different from the value we had in our previous study \cite{Geng:2021BHl} of $\theta_P \approx 0.975$.

\subsection{Page angles}

\begin{figure}
    \centering
    \includegraphics[width=0.75\linewidth]{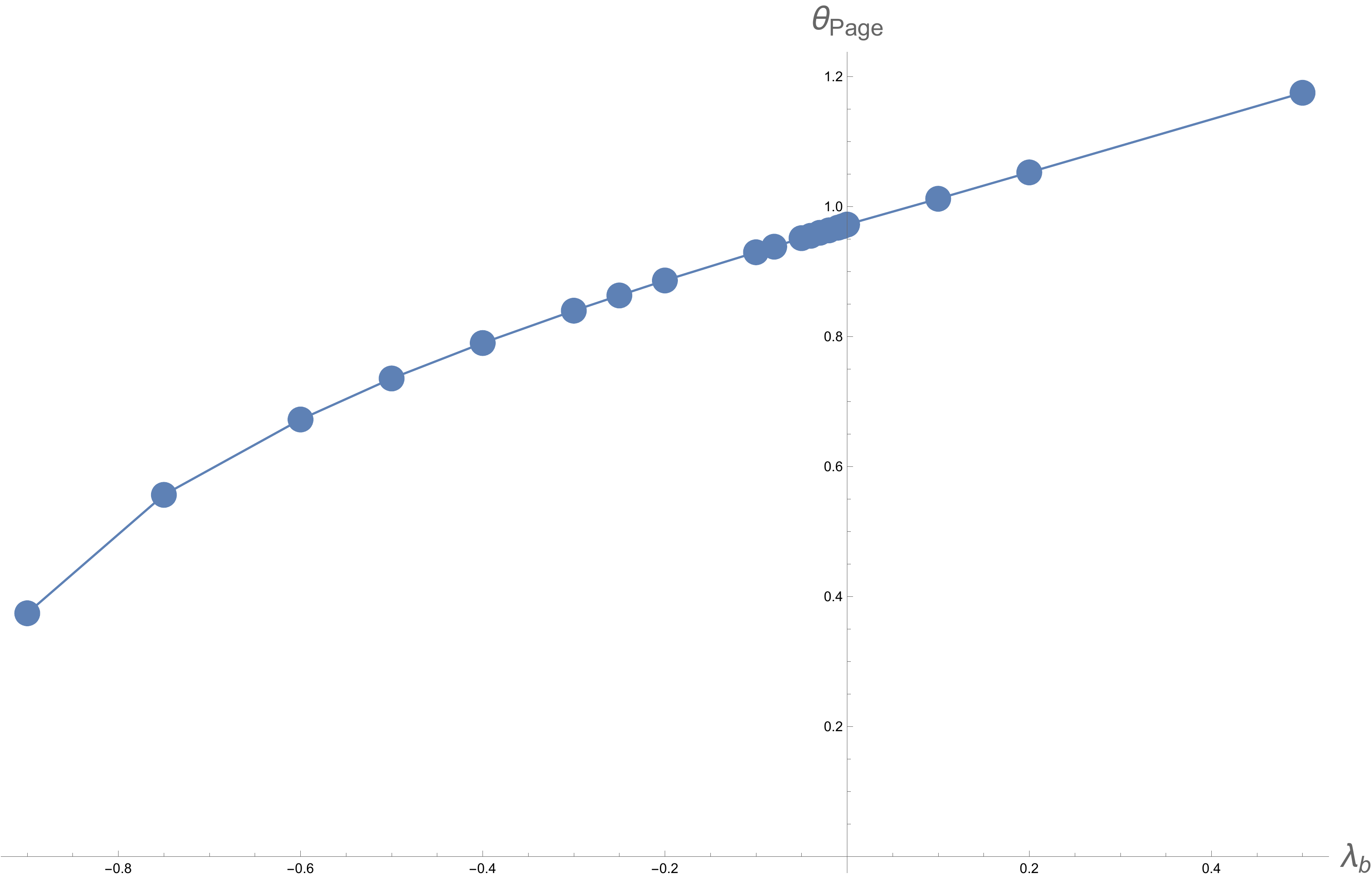}
    \caption{Page angles $\theta_{Page}$ as a function of $\lambda_b$. At $\lambda_b=0$, we obtain $\theta_P\approx 0.972$, matching our previous results. Furthermore the curve monotonically increases.}
    \label{fig:Pageangles}
\end{figure}

Extracting the Page angle, namely the angle for which the area of the curve shooting to the defect vanishes, for each value of $\lambda_b$ in figure \ref{fig:PhaseStructure} we obtain the figure \ref{fig:Pageangles}.

As we noted before, we can check our method by comparing the $\theta_P$ we obtain now compared to what we obtained before \cite{Geng:2021BHl} and we observe that this is roughly the same, namely approximately $0.972$.

We can see that the Page angle monotonically increases with the value of $\lambda_b$. Therefore, there is another interesting quantity we can extract from this plot and this is the value of the DGP coupling for which the Page angle is the critical angle. We recall that in empty $AdS$ the Page angle is the same as the critical angle. The reason why this is not true for the black string, is because the HM has a non-zero renormalised area and hence subtracts a small number from the island surface area at the critical angle. Thus, the area at the critical angle is slightly negative and hence the value of $\theta_b$ for which the area vanishes lies slightly below $\theta_c$. In the case at hand, by fine tuning the strength of the gravity in the brane, namely the DGP coupling parameter, $\lambda_b$, we can lift the effect of the black string and make the Page angle match the critical angle again. A similar thing has been done in \cite{Merna}, where the fine tuned parameter is the black hole size. This happens for approximately $\lambda_b \approx 0.0364$.

%%%%%%%%%%%%%%%%%
\section{Conclusions} \label{Section5}
%%%%%%%%%%%%%%%%%

We have extended the study in \cite{Geng:2021BHl} analysing the subregion entropy for a doubly holographic black string model. The system is dual to a $BCFT_d$ with a black-hole background and to a non-gravitating bath coupled to an $AdS_d$ brane in which dynamical gravity can been turned on. This dynamical gravity is parameterised by a coupling $\lambda_b$ which measures the strength of gravity on the brane. In this way we are testing the parameter space of the black string model with respect to the anchoring point on the bath, the $AdS_d$ brane angle and the strength of the DGP term, $(\Gamma, \theta_b, \lambda_b)$.

In empty $AdS$ we previously observed that island surfaces did not exist below the critical angle. This issue was resolved at finite temperature, namely for the $\lambda_b=0$ system giving us a rich phase structure. However this phase structure could be modified by the inclusion of dynamical gravity. Therefore our aim here was to limit the parameter space for well-behaved theories supporting islands and to study the subregion entropy phases when the strength of gravity on the brane was modified. In particular, throughout this note we focused on unitary evolution of entropy and the existence of a Page curve.

In order to do this, we have first proceeded to study the RT structure of the system, namely the existence and number of competing RT-candidates. For most of the parameter space there was a single island RT-candidate for a shooting point on the brane. This shooting point was limited to only a fraction of the brane, characterising the size of the island. This extends our critical anchor story, since now, changing the $\lambda_b$ parameter allows for non-zero critical anchors even above the critical angle. For the particular case of $\lambda_b=-1$ we observed 3 competing RT-candidates and steep phase transition between two of these. This phase transition was captured by a special value of anchoring point on the bath $\Gamma_{min}$. All of this shows a richer RT structure which we needed to take into account when computing their areas.

With this in mind we computed the areas and presented them as 2D surfaces. Some of the behaviour they exhibited matched our previous work but some of it was new and needed to be explained. In particular, as mentioned, we highlight the case of $\lambda=-1$ and how the tiny island effects were lifted by large absolute value strength of DGP coupling. All of this is summarised in figure \ref{fig:PhaseStructure}, which constitutes the main result in this note. Here we distinguish the parts of parameter space which give us a Page curve from those that give us an eternally constant entropy. In any case it seems both situations present a unitary evolution and hence means that the theories in the parameter space are stable against unitarity tests.

We can use this information to obtain the value of $\lambda_b$ for which the critical angle matches the Page angle, which we obtained to be $\lambda_b \approx 0.0364$. In other words, one can fine tune the strength of gravity on the brane so that the effect of the black string on the renormalised area of HM is lifted and we obtain what we had in the case of empty $AdS$. The difference is that now, unlike in empty $AdS$ we do have island surfaces below the critical angle.

In recent work it was discussed that the DGP terms may solve the issue of massive gravity in island models \cite{asRong-Xin_Miao1,Rong-Xin_Miao2}. In this note we have seen how the parameter space is confined in order for islands to exist. Something similar might be expected for wedge holographic models were we can also impose other constraints. This is going to be analysed in more detail in future work \cite{GengDGP}.

This study has focused on the entropy phase structure and RT surfaces for angles ${\theta_b< \pi/2}$. However, the relationship between brane tension and angle indicates that for larger $\lambda_b$, there may be a range of subcritical brane angles that are still physical above $\pi/2$. It would be interesting to explore this further for larger angles and values of DGP coupling, and determine how the competition between three RT surfaces is resolved and if any phase transitions occur.

\section*{Acknowledgments}

I am grateful to Hao Geng, Andreas Karch, Suvrat Raju, Lisa Randall, Marcos Riojas, Sanjit Shashi and Merna Youssef for useful discussions and collaborations which led to this work. I would also like to thank Elena Cáceres, Rodrigo Castillo Vásquez, Hao Geng, Andreas Karch, Marcos Riojas and Sanjit Shashi for comments on the draft. This work is supported in part by the National Science Foundation under Grant No.~PHY-1914679 and by the Robert N. Little Fellowship.

\begin{appendices}

%%%%%%%%%%%%%
\section{Supporting Plots}\label{app:A}
%%%%%%%%%%%%%
Here we show some of the plots which were used in part of the computations in the main text. They illustrate some of the descriptions and claims above.

\begin{figure}[h]
    \centering
    \includegraphics[width=0.7\linewidth]{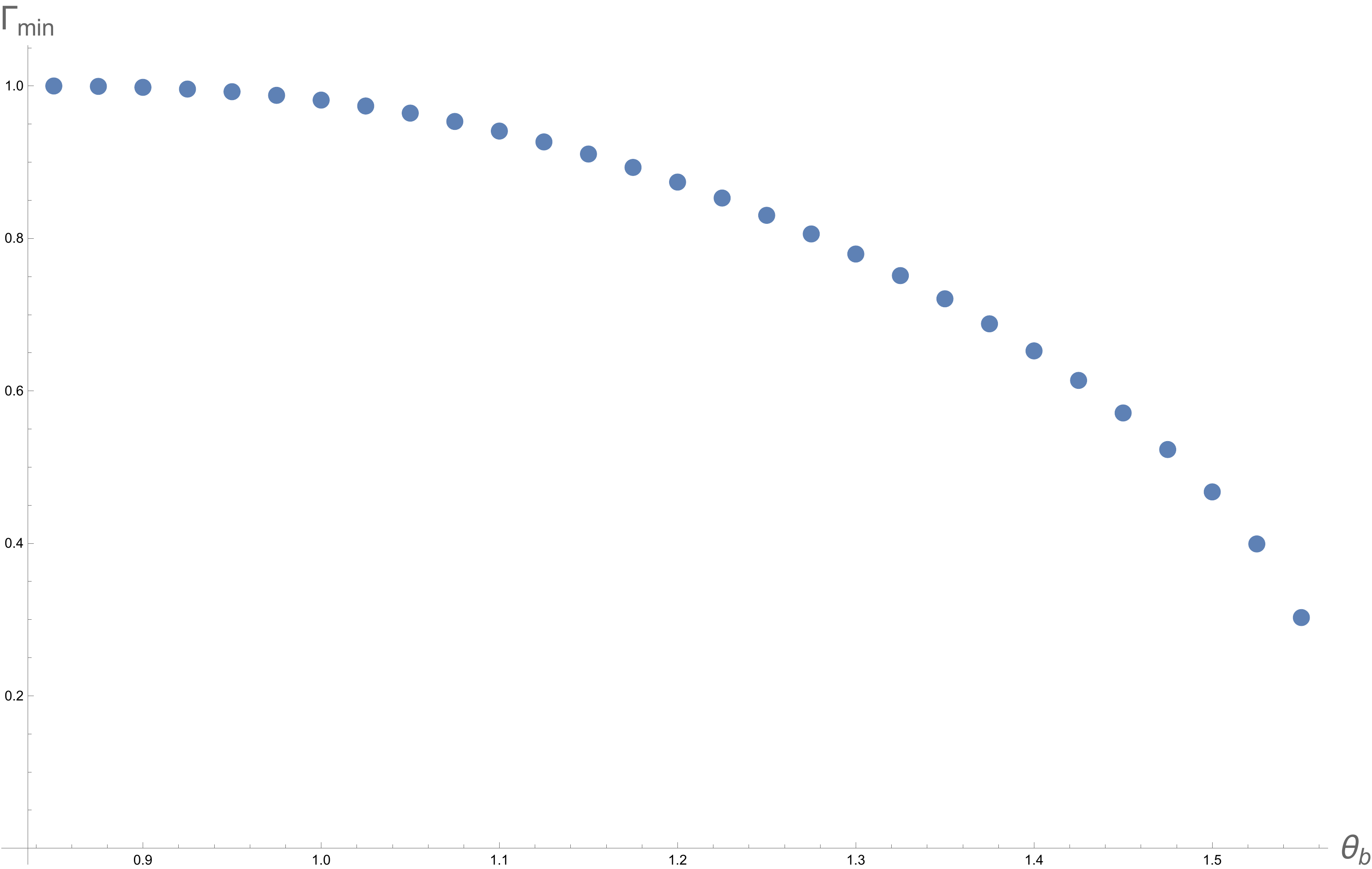}
    \caption{Plot of $\Gamma_{min}$ for $\lambda_b=-1$.}
    \label{fig:Supp1}
\end{figure} 

Figure \ref{fig:Supp1} shows what we claimed for $\lambda_b=-1$. This is that $\Gamma_{min}$ decreases as we increase the brane angle.

\begin{figure}[h]
    \centering
    \includegraphics[width=0.75\linewidth]{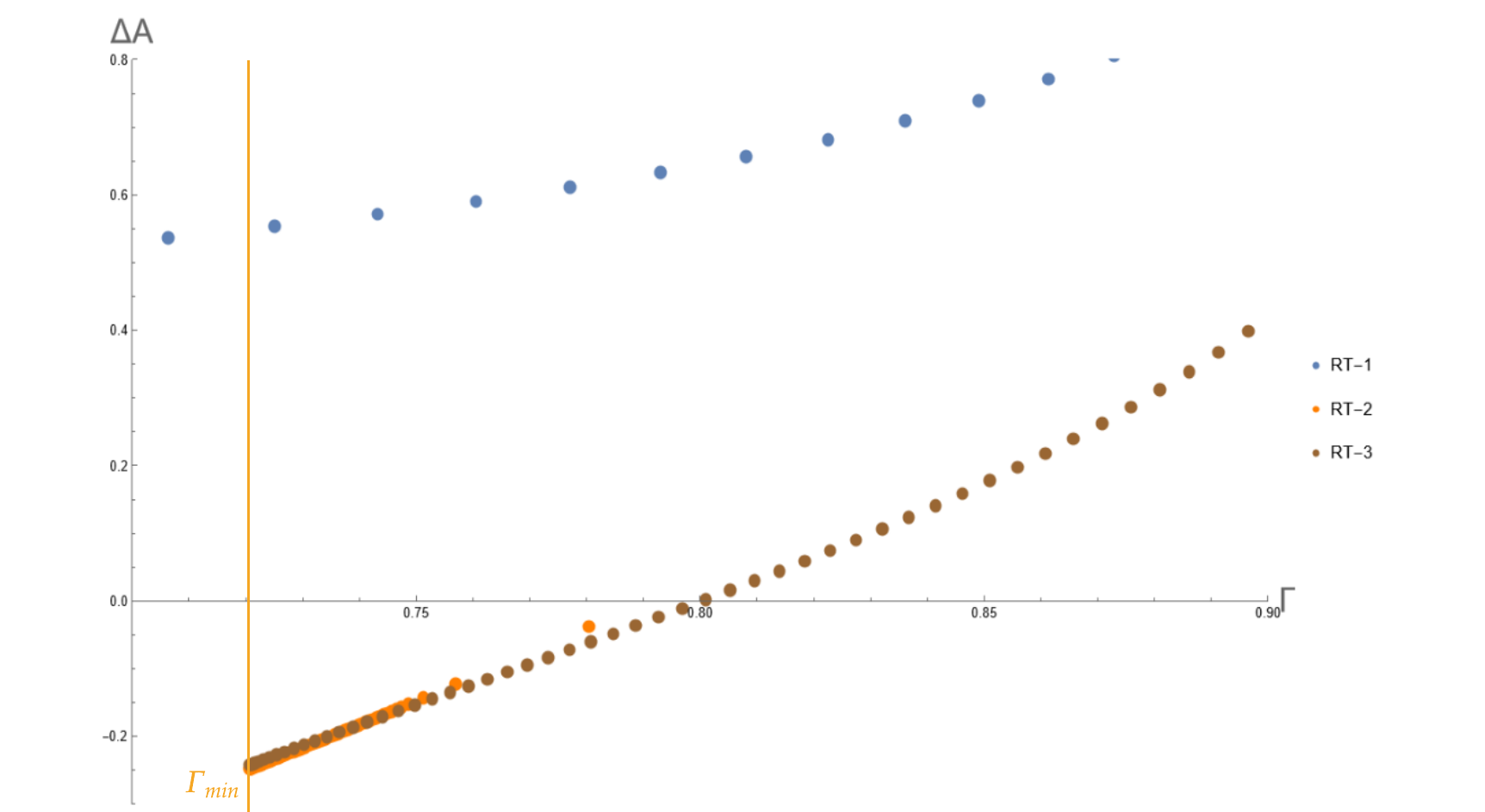}
    \caption{Areas for RT-1, RT-2 and RT-3 for $\lambda_b=-1$ and $\theta_b=1.35$. The yellow line marks the position of $\Gamma_{min}$.}
    \label{fig:Supp2}
\end{figure}

Figure \ref{fig:Supp2} shows the areas for two of the competing surfaces when $\lambda_b=-1$ and $\theta_b=1.35$. The surfaces RT-2 and RT-3 become degenerate at $\Gamma_{min}$, while the surfaces RT-1 and RT-2 become degenerate at $\Gamma=1$. Above this value we have three competing RT surfaces: RT-1, RT-2 and RT-3. We can see how RT-2 and RT-3 have increasing area as $\Gamma \rightarrow 1$, but overall, RT-3 has a slightly smaller area than RT-2 

\end{appendices}

\bibliographystyle{JHEP}
\bibliography{references}

\providecommand{\href}[2]{#2}\begingroup\raggedright\begin{thebibliography}{100}

\bibitem{Ryu:2006bv}
S.~Ryu and T.~Takayanagi, \emph{{Holographic derivation of entanglement entropy
  from AdS/CFT}},
  \href{https://doi.org/10.1103/PhysRevLett.96.181602}{\emph{Phys. Rev. Lett.}
  {\bfseries 96} (2006) 181602}
  [\href{https://arxiv.org/abs/hep-th/0603001}{{\ttfamily hep-th/0603001}}].

\bibitem{Ryu:2006ef}
S.~Ryu and T.~Takayanagi, \emph{{Aspects of Holographic Entanglement Entropy}},
  \href{https://doi.org/10.1088/1126-6708/2006/08/045}{\emph{JHEP} {\bfseries
  08} (2006) 045} [\href{https://arxiv.org/abs/hep-th/0605073}{{\ttfamily
  hep-th/0605073}}].

\bibitem{Hubeny:2007xt}
V.~E. Hubeny, M.~Rangamani and T.~Takayanagi, \emph{{A Covariant holographic
  entanglement entropy proposal}},
  \href{https://doi.org/10.1088/1126-6708/2007/07/062}{\emph{JHEP} {\bfseries
  07} (2007) 062} [\href{https://arxiv.org/abs/0705.0016}{{\ttfamily
  0705.0016}}].

\bibitem{Faulkner:2013ana}
T.~Faulkner, A.~Lewkowycz and J.~Maldacena, \emph{{Quantum corrections to
  holographic entanglement entropy}},
  \href{https://doi.org/10.1007/JHEP11(2013)074}{\emph{JHEP} {\bfseries 11}
  (2013) 074} [\href{https://arxiv.org/abs/1307.2892}{{\ttfamily 1307.2892}}].

\bibitem{Lewkowycz:2013nqa}
A.~Lewkowycz and J.~Maldacena, \emph{{Generalized gravitational entropy}},
  \href{https://doi.org/10.1007/JHEP08(2013)090}{\emph{JHEP} {\bfseries 08}
  (2013) 090} [\href{https://arxiv.org/abs/1304.4926}{{\ttfamily 1304.4926}}].

\bibitem{Engelhardt:2014gca}
N.~Engelhardt and A.~C. Wall, \emph{{Quantum Extremal Surfaces: Holographic
  Entanglement Entropy beyond the Classical Regime}},
  \href{https://doi.org/10.1007/JHEP01(2015)073}{\emph{JHEP} {\bfseries 01}
  (2015) 073} [\href{https://arxiv.org/abs/1408.3203}{{\ttfamily 1408.3203}}].

\bibitem{Maldacena:1997re}
J.~M. Maldacena, \emph{{The Large N limit of superconformal field theories and
  supergravity}}, \href{https://doi.org/10.1023/A:1026654312961}{\emph{Int. J.
  Theor. Phys.} {\bfseries 38} (1999) 1113}
  [\href{https://arxiv.org/abs/hep-th/9711200}{{\ttfamily hep-th/9711200}}].

\bibitem{Gubser:1998bc}
S.~Gubser, I.~R. Klebanov and A.~M. Polyakov, \emph{{Gauge theory correlators
  from noncritical string theory}},
  \href{https://doi.org/10.1016/S0370-2693(98)00377-3}{\emph{Phys. Lett. B}
  {\bfseries 428} (1998) 105}
  [\href{https://arxiv.org/abs/hep-th/9802109}{{\ttfamily hep-th/9802109}}].

\bibitem{Witten:1998qj}
E.~Witten, \emph{{Anti-de Sitter space and holography}},
  \href{https://doi.org/10.4310/ATMP.1998.v2.n2.a2}{\emph{Adv. Theor. Math.
  Phys.} {\bfseries 2} (1998) 253}
  [\href{https://arxiv.org/abs/hep-th/9802150}{{\ttfamily hep-th/9802150}}].

\bibitem{Cardy:2004hm}
J.~L. Cardy, \emph{{Boundary conformal field theory}},
  \href{https://arxiv.org/abs/hep-th/0411189}{{\ttfamily hep-th/0411189}}.

\bibitem{Karch:2000gx}
A.~Karch and L.~Randall, \emph{{Open and closed string interpretation of SUSY
  CFT's on branes with boundaries}},
  \href{https://doi.org/10.1088/1126-6708/2001/06/063}{\emph{JHEP} {\bfseries
  06} (2001) 063} [\href{https://arxiv.org/abs/hep-th/0105132}{{\ttfamily
  hep-th/0105132}}].

\bibitem{Takayanagi:2011zk}
T.~Takayanagi, \emph{{Holographic Dual of BCFT}},
  \href{https://doi.org/10.1103/PhysRevLett.107.101602}{\emph{Phys. Rev. Lett.}
  {\bfseries 107} (2011) 101602}
  [\href{https://arxiv.org/abs/1105.5165}{{\ttfamily 1105.5165}}].

\bibitem{Fujita:2011fp}
M.~Fujita, T.~Takayanagi and E.~Tonni, \emph{{Aspects of AdS/BCFT}},
  \href{https://doi.org/10.1007/JHEP11(2011)043}{\emph{JHEP} {\bfseries 11}
  (2011) 043} [\href{https://arxiv.org/abs/1108.5152}{{\ttfamily 1108.5152}}].

\bibitem{Karch:2000ct}
A.~Karch and L.~Randall, \emph{{Locally localized gravity}},
  \href{https://doi.org/10.1088/1126-6708/2001/05/008}{\emph{JHEP} {\bfseries
  05} (2001) 008} [\href{https://arxiv.org/abs/hep-th/0011156}{{\ttfamily
  hep-th/0011156}}].

\bibitem{Randall:1999vf}
L.~Randall and R.~Sundrum, \emph{{An Alternative to compactification}},
  \href{https://doi.org/10.1103/PhysRevLett.83.4690}{\emph{Phys. Rev. Lett.}
  {\bfseries 83} (1999) 4690}
  [\href{https://arxiv.org/abs/hep-th/9906064}{{\ttfamily hep-th/9906064}}].

\bibitem{Penington:2019npb}
G.~Penington, \emph{{Entanglement Wedge Reconstruction and the Information
  Paradox}}, \href{https://doi.org/10.1007/JHEP09(2020)002}{\emph{JHEP}
  {\bfseries 09} (2020) 002}
  [\href{https://arxiv.org/abs/1905.08255}{{\ttfamily 1905.08255}}].

\bibitem{Almheiri:2019psf}
A.~Almheiri, N.~Engelhardt, D.~Marolf and H.~Maxfield, \emph{{The entropy of
  bulk quantum fields and the entanglement wedge of an evaporating black
  hole}}, \href{https://doi.org/10.1007/JHEP12(2019)063}{\emph{JHEP} {\bfseries
  12} (2019) 063} [\href{https://arxiv.org/abs/1905.08762}{{\ttfamily
  1905.08762}}].

\bibitem{Almheiri:2019hni}
A.~Almheiri, R.~Mahajan, J.~Maldacena and Y.~Zhao, \emph{{The Page curve of
  Hawking radiation from semiclassical geometry}},
  \href{https://doi.org/10.1007/JHEP03(2020)149}{\emph{JHEP} {\bfseries 03}
  (2020) 149} [\href{https://arxiv.org/abs/1908.10996}{{\ttfamily
  1908.10996}}].

\bibitem{Almheiri:2019psy}
A.~Almheiri, R.~Mahajan and J.~E. Santos, \emph{{Entanglement islands in higher
  dimensions}},
  \href{https://doi.org/10.21468/SciPostPhys.9.1.001}{\emph{SciPost Phys.}
  {\bfseries 9} (2020) 001} [\href{https://arxiv.org/abs/1911.09666}{{\ttfamily
  1911.09666}}].

\bibitem{Almheiri:2019yqk}
A.~Almheiri, R.~Mahajan and J.~Maldacena, \emph{{Islands outside the horizon}},
   \href{https://arxiv.org/abs/1910.11077}{{\ttfamily 1910.11077}}.

\bibitem{Geng:2021BHl}
H.~Geng, A.~Karch, C.~Perez-Pardavila, S.~Raju, L.~Randall, M.~Riojas and
  S.~Shashi, \emph{{Entanglement phase structure of a holographic BCFT in a
  black hole background}},
  \href{https://doi.org/10.1007/JHEP05(2022)153}{\emph{JHEP} {\bfseries 05}
  (2022) 153} [\href{https://arxiv.org/abs/2112.09132}{{\ttfamily
  2112.09132}}].

\bibitem{Geng:2020fxl}
H.~Geng, A.~Karch, C.~Perez-Pardavila, S.~Raju, L.~Randall, M.~Riojas and
  S.~Shashi, \emph{{Information Transfer with a Gravitating Bath}},
  \href{https://arxiv.org/abs/2012.04671}{{\ttfamily 2012.04671}}.

\bibitem{lubkin1978entropy}
E.~Lubkin, \emph{Entropy of an n-system from its correlation with a
  k-reservoir}, \href{https://doi.org/10.1063/1.523763}{\emph{Journal of
  Mathematical Physics} {\bfseries 19} (1978) 1028}.

\bibitem{Page:1993df}
D.~N. Page, \emph{{Average entropy of a subsystem}},
  \href{https://doi.org/10.1103/PhysRevLett.71.1291}{\emph{Phys. Rev. Lett.}
  {\bfseries 71} (1993) 1291}
  [\href{https://arxiv.org/abs/gr-qc/9305007}{{\ttfamily gr-qc/9305007}}].

\bibitem{Hartman:2013qma}
T.~Hartman and J.~Maldacena, \emph{{Time Evolution of Entanglement Entropy from
  Black Hole Interiors}},
  \href{https://doi.org/10.1007/JHEP05(2013)014}{\emph{JHEP} {\bfseries 05}
  (2013) 014} [\href{https://arxiv.org/abs/1303.1080}{{\ttfamily 1303.1080}}].

\bibitem{Ling:2020laa}
Y.~Ling, Y.~Liu and Z.-Y. Xian, \emph{{Island in Charged Black Holes}},
  \href{https://doi.org/10.1007/JHEP03(2021)251}{\emph{JHEP} {\bfseries 03}
  (2021) 251} [\href{https://arxiv.org/abs/2010.00037}{{\ttfamily
  2010.00037}}].

\bibitem{KumarBasak:2020ams}
J.~Kumar~Basak, D.~Basu, V.~Malvimat, H.~Parihar and G.~Sengupta,
  \emph{{Islands for Entanglement Negativity}},
  \href{https://arxiv.org/abs/2012.03983}{{\ttfamily 2012.03983}}.

\bibitem{Emparan:2020znc}
R.~Emparan, A.~M. Frassino and B.~Way, \emph{{Quantum BTZ black hole}},
  \href{https://doi.org/10.1007/JHEP11(2020)137}{\emph{JHEP} {\bfseries 11}
  (2020) 137} [\href{https://arxiv.org/abs/2007.15999}{{\ttfamily
  2007.15999}}].

\bibitem{Caceres:2020jcn}
E.~Caceres, A.~Kundu, A.~K. Patra and S.~Shashi, \emph{{Warped Information and
  Entanglement Islands in AdS/WCFT}},
  \href{https://arxiv.org/abs/2012.05425}{{\ttfamily 2012.05425}}.

\bibitem{Caceres:2021fuw}
E.~Caceres, A.~Kundu, A.~K. Patra and S.~Shashi, \emph{{Page Curves and Bath
  Deformations}},  \href{https://arxiv.org/abs/2107.00022}{{\ttfamily
  2107.00022}}.

\bibitem{Deng:2020ent}
F.~Deng, J.~Chu and Y.~Zhou, \emph{{Defect extremal surface as the holographic
  counterpart of Island formula}},
  \href{https://doi.org/10.1007/JHEP03(2021)008}{\emph{JHEP} {\bfseries 03}
  (2021) 008} [\href{https://arxiv.org/abs/2012.07612}{{\ttfamily
  2012.07612}}].

\bibitem{Krishnan:2020fer}
C.~Krishnan, \emph{{Critical Islands}},
  \href{https://doi.org/10.1007/JHEP01(2021)179}{\emph{JHEP} {\bfseries 01}
  (2021) 179} [\href{https://arxiv.org/abs/2007.06551}{{\ttfamily
  2007.06551}}].

\bibitem{Balasubramanian:2020coy}
V.~Balasubramanian, A.~Kar and T.~Ugajin, \emph{{Entanglement between two
  disjoint universes}},
  \href{https://doi.org/10.1007/JHEP02(2021)136}{\emph{JHEP} {\bfseries 02}
  (2021) 136} [\href{https://arxiv.org/abs/2008.05274}{{\ttfamily
  2008.05274}}].

\bibitem{Balasubramanian:2020xqf}
V.~Balasubramanian, A.~Kar and T.~Ugajin, \emph{{Islands in de Sitter space}},
  \href{https://doi.org/10.1007/JHEP02(2021)072}{\emph{JHEP} {\bfseries 02}
  (2021) 072} [\href{https://arxiv.org/abs/2008.05275}{{\ttfamily
  2008.05275}}].

\bibitem{Manu:2020tty}
A.~Manu, K.~Narayan and P.~Paul, \emph{{Cosmological singularities,
  entanglement and quantum extremal surfaces}},
  \href{https://doi.org/10.1007/JHEP04(2021)200}{\emph{JHEP} {\bfseries 04}
  (2021) 200} [\href{https://arxiv.org/abs/2012.07351}{{\ttfamily
  2012.07351}}].

\bibitem{Karlsson:2021vlh}
A.~Karlsson, \emph{{Concerns about the replica wormhole derivation of the
  island conjecture}},  \href{https://arxiv.org/abs/2101.05879}{{\ttfamily
  2101.05879}}.

\bibitem{Wang:2021woy}
X.~Wang, R.~Li and J.~Wang, \emph{{Islands and Page curves of
  Reissner-Nordstr\"om black holes}},
  \href{https://doi.org/10.1007/JHEP04(2021)103}{\emph{JHEP} {\bfseries 04}
  (2021) 103} [\href{https://arxiv.org/abs/2101.06867}{{\ttfamily
  2101.06867}}].

\bibitem{Miao:2021ual}
R.-X. Miao, \emph{{Codimension-n Holography for the Cones}},
  \href{https://arxiv.org/abs/2101.10031}{{\ttfamily 2101.10031}}.

\bibitem{Bachas:2021fqo}
C.~Bachas and V.~Papadopoulos, \emph{{Phases of Holographic Interfaces}},
  \href{https://doi.org/10.1007/JHEP04(2021)262}{\emph{JHEP} {\bfseries 04}
  (2021) 262} [\href{https://arxiv.org/abs/2101.12529}{{\ttfamily
  2101.12529}}].

\bibitem{May:2021zyu}
A.~May and D.~Wakeham, \emph{{Quantum tasks require islands on the brane}},
  \href{https://doi.org/10.1088/1361-6382/ac025d}{\emph{Class. Quant. Grav.}
  {\bfseries 38} (2021) 144001}
  [\href{https://arxiv.org/abs/2102.01810}{{\ttfamily 2102.01810}}].

\bibitem{Kawabata:2021hac}
K.~Kawabata, T.~Nishioka, Y.~Okuyama and K.~Watanabe, \emph{{Probing Hawking
  radiation through capacity of entanglement}},
  \href{https://doi.org/10.1007/JHEP05(2021)062}{\emph{JHEP} {\bfseries 05}
  (2021) 062} [\href{https://arxiv.org/abs/2102.02425}{{\ttfamily
  2102.02425}}].

\bibitem{Bhattacharya:2021jrn}
A.~Bhattacharya, A.~Bhattacharyya, P.~Nandy and A.~K. Patra, \emph{{Islands and
  complexity of eternal black hole and radiation subsystems for a doubly
  holographic model}},
  \href{https://doi.org/10.1007/JHEP05(2021)135}{\emph{JHEP} {\bfseries 05}
  (2021) 135} [\href{https://arxiv.org/abs/2103.15852}{{\ttfamily
  2103.15852}}].

\bibitem{Anderson:2021vof}
L.~Anderson, O.~Parrikar and R.~M. Soni, \emph{{Islands with Gravitating
  Baths}},  \href{https://arxiv.org/abs/2103.14746}{{\ttfamily 2103.14746}}.

\bibitem{Miyata:2021ncm}
A.~Miyata and T.~Ugajin, \emph{{Evaporation of black holes in flat space
  entangled with an auxiliary universe}},
  \href{https://arxiv.org/abs/2104.00183}{{\ttfamily 2104.00183}}.

\bibitem{Kim:2021gzd}
W.~Kim and M.~Nam, \emph{{Entanglement entropy of asymptotically flat
  non-extremal and extremal black holes with an island}},
  \href{https://arxiv.org/abs/2103.16163}{{\ttfamily 2103.16163}}.

\bibitem{Hollowood:2021nlo}
T.~J. Hollowood, S.~Prem~Kumar, A.~Legramandi and N.~Talwar, \emph{{Islands in
  the Stream of Hawking Radiation}},
  \href{https://arxiv.org/abs/2104.00052}{{\ttfamily 2104.00052}}.

\bibitem{Wang:2021mqq}
X.~Wang, R.~Li and J.~Wang, \emph{{Page curves for a family of exactly solvable
  evaporating black holes}},
  \href{https://doi.org/10.1103/PhysRevD.103.126026}{\emph{Phys. Rev. D}
  {\bfseries 103} (2021) 126026}
  [\href{https://arxiv.org/abs/2104.00224}{{\ttfamily 2104.00224}}].

\bibitem{Aalsma:2021bit}
L.~Aalsma and W.~Sybesma, \emph{{The Price of Curiosity: Information Recovery
  in de Sitter Space}},
  \href{https://doi.org/10.1007/JHEP05(2021)291}{\emph{JHEP} {\bfseries 05}
  (2021) 291} [\href{https://arxiv.org/abs/2104.00006}{{\ttfamily
  2104.00006}}].

\bibitem{Ghosh:2021axl}
K.~Ghosh and C.~Krishnan, \emph{{Dirichlet Baths and the Not-so-Fine-Grained
  Page Curve}},  \href{https://arxiv.org/abs/2103.17253}{{\ttfamily
  2103.17253}}.

\bibitem{Neuenfeld:2021wbl}
D.~Neuenfeld, \emph{{The Dictionary for Double Holography and Graviton Masses
  in d Dimensions}},  \href{https://arxiv.org/abs/2104.02801}{{\ttfamily
  2104.02801}}.

\bibitem{Geng:2021iyq}
H.~Geng, S.~L\"ust, R.~K. Mishra and D.~Wakeham, \emph{{Holographic BCFTs and
  Communicating Black Holes}},
  \href{https://arxiv.org/abs/2104.07039}{{\ttfamily 2104.07039}}.

\bibitem{Balasubramanian:2021wgd}
V.~Balasubramanian, A.~Kar and T.~Ugajin, \emph{{Entanglement between two
  gravitating universes}},  \href{https://arxiv.org/abs/2104.13383}{{\ttfamily
  2104.13383}}.

\bibitem{Uhlemann:2021nhu}
C.~F. Uhlemann, \emph{{Islands and Page curves in 4d from Type IIB}},
  \href{https://arxiv.org/abs/2105.00008}{{\ttfamily 2105.00008}}.

\bibitem{Neuenfeld:2021bsb}
D.~Neuenfeld, \emph{{Double Holography as a Model for Black Hole
  Complementarity}},  \href{https://arxiv.org/abs/2105.01130}{{\ttfamily
  2105.01130}}.

\bibitem{Kawabata:2021vyo}
K.~Kawabata, T.~Nishioka, Y.~Okuyama and K.~Watanabe, \emph{{Replica wormholes
  and capacity of entanglement}},
  \href{https://arxiv.org/abs/2105.08396}{{\ttfamily 2105.08396}}.

\bibitem{Chu:2021gdb}
J.~Chu, F.~Deng and Y.~Zhou, \emph{{Page Curve from Defect Extremal Surface and
  Island in Higher Dimensions}},
  \href{https://arxiv.org/abs/2105.09106}{{\ttfamily 2105.09106}}.

\bibitem{Kruthoff:2021vgv}
J.~Kruthoff, R.~Mahajan and C.~Murdia, \emph{{Free fermion entanglement with a
  semitransparent interface: the effect of graybody factors on entanglement
  islands}},  \href{https://arxiv.org/abs/2106.10287}{{\ttfamily 2106.10287}}.

\bibitem{Akal:2021foz}
I.~Akal, Y.~Kusuki, N.~Shiba, T.~Takayanagi and Z.~Wei, \emph{{Holographic
  moving mirrors}},  \href{https://arxiv.org/abs/2106.11179}{{\ttfamily
  2106.11179}}.

\bibitem{KumarBasak:2021rrx}
J.~Kumar~Basak, D.~Basu, V.~Malvimat, H.~Parihar and G.~Sengupta, \emph{{Page
  Curve for Entanglement Negativity through Geometric Evaporation}},
  \href{https://arxiv.org/abs/2106.12593}{{\ttfamily 2106.12593}}.

\bibitem{Lu:2021gmv}
Y.~Lu and J.~Lin, \emph{{Islands in Kaluza-Klein black holes}},
  \href{https://arxiv.org/abs/2106.07845}{{\ttfamily 2106.07845}}.

\bibitem{Omiya:2021olc}
H.~Omiya and Z.~Wei, \emph{{Causal Structures and Nonlocality in Double
  Holography}},  \href{https://arxiv.org/abs/2107.01219}{{\ttfamily
  2107.01219}}.

\bibitem{Ahn:2021chg}
B.~Ahn, S.-E. Bak, H.-S. Jeong, K.-Y. Kim and Y.-W. Sun, \emph{{Islands in
  charged linear dilaton black holes}},
  \href{https://arxiv.org/abs/2107.07444}{{\ttfamily 2107.07444}}.

\bibitem{Balasubramanian:2021xcm}
V.~Balasubramanian, B.~Craps, M.~Khramtsov and E.~Shaghoulian,
  \emph{{Submerging islands through thermalization}},
  \href{https://doi.org/10.1007/JHEP10(2021)048}{\emph{JHEP} {\bfseries 10}
  (2021) 048} [\href{https://arxiv.org/abs/2107.14746}{{\ttfamily
  2107.14746}}].

\bibitem{Li:2021dmf}
T.~Li, M.-K. Yuan and Y.~Zhou, \emph{{Defect Extremal Surface for Reflected
  Entropy}},  \href{https://arxiv.org/abs/2108.08544}{{\ttfamily 2108.08544}}.

\bibitem{Kames-King:2021etp}
J.~Kames-King, E.~Verheijden and E.~Verlinde, \emph{{No Page Curves for the de
  Sitter Horizon}},  \href{https://arxiv.org/abs/2108.09318}{{\ttfamily
  2108.09318}}.

\bibitem{Sun:2021dfl}
P.-C. Sun, \emph{{Entanglement Islands from Holographic Thermalization of
  Rotating Charged Black Hole}},
  \href{https://arxiv.org/abs/2108.12557}{{\ttfamily 2108.12557}}.

\bibitem{Hollowood:2021wkw}
T.~J. Hollowood, S.~P. Kumar, A.~Legramandi and N.~Talwar, \emph{{Ephemeral
  Islands, Plunging Quantum Extremal Surfaces and BCFT channels}},
  \href{https://arxiv.org/abs/2109.01895}{{\ttfamily 2109.01895}}.

\bibitem{Miyaji:2021lcq}
M.~Miyaji, \emph{{Island for Gravitationally Prepared State and Pseudo
  Entanglement Wedge}},  \href{https://arxiv.org/abs/2109.03830}{{\ttfamily
  2109.03830}}.

\bibitem{Bhattacharya:2021dnd}
A.~Bhattacharya, A.~Bhattacharyya, P.~Nandy and A.~K. Patra, \emph{{Partial
  islands and subregion complexity in geometric secret-sharing model}},
  \href{https://arxiv.org/abs/2109.07842}{{\ttfamily 2109.07842}}.

\bibitem{Goswami:2021ksw}
K.~Goswami, K.~Narayan and H.~K. Saini, \emph{{Cosmologies, singularities and
  quantum extremal surfaces}},
  \href{https://arxiv.org/abs/2111.14906}{{\ttfamily 2111.14906}}.

\bibitem{Chu:2021mvq}
C.-S. Chu and R.-X. Miao, \emph{{Conformal Boundary Condition and Perturbation
  Spectrum in AdS/BCFT}},  \href{https://arxiv.org/abs/2110.03159}{{\ttfamily
  2110.03159}}.

\bibitem{Arefeva:2021kfx}
I.~Aref'eva and I.~Volovich, \emph{{A Note on Islands in Schwarzschild Black
  Holes}},  \href{https://arxiv.org/abs/2110.04233}{{\ttfamily 2110.04233}}.

\bibitem{Shaghoulian:2021cef}
E.~Shaghoulian, \emph{{The central dogma and cosmological horizons}},
  \href{https://arxiv.org/abs/2110.13210}{{\ttfamily 2110.13210}}.

\bibitem{Garcia-Garcia:2021squ}
A.~M. Garc\'\i{}a-Garc\'\i{}a and V.~Godet, \emph{{Half-wormholes in nearly
  AdS$_2$ holography}},  \href{https://arxiv.org/abs/2107.07720}{{\ttfamily
  2107.07720}}.

\bibitem{Buoninfante:2021ijy}
L.~Buoninfante, F.~Di~Filippo and S.~Mukohyama, \emph{{On the assumptions
  leading to the information loss paradox}},
  \href{https://arxiv.org/abs/2107.05662}{{\ttfamily 2107.05662}}.

\bibitem{Yu:2021cgi}
M.-H. Yu and X.-H. Ge, \emph{{Page Curves and Islands in Charged Dilaton Black
  Holes}},  \href{https://arxiv.org/abs/2107.03031}{{\ttfamily 2107.03031}}.

\bibitem{Nam:2021bml}
C.~H. Nam, \emph{{Entanglement entropy and Page curve of black holes with
  island in massive gravity}},
  \href{https://arxiv.org/abs/2108.10144}{{\ttfamily 2108.10144}}.

\bibitem{He:2021mst}
S.~He, Y.~Sun, L.~Zhao and Y.-X. Zhang, \emph{{The universality of islands
  outside the horizon}},  \href{https://arxiv.org/abs/2110.07598}{{\ttfamily
  2110.07598}}.

\bibitem{Langhoff:2021uct}
K.~Langhoff, C.~Murdia and Y.~Nomura, \emph{{Multiverse in an inverted
  island}}, \href{https://doi.org/10.1103/PhysRevD.104.086007}{\emph{Phys. Rev.
  D} {\bfseries 104} (2021) 086007}
  [\href{https://arxiv.org/abs/2106.05271}{{\ttfamily 2106.05271}}].

\bibitem{Ageev:2021ipd}
D.~S. Ageev, \emph{{Shaping contours of entanglement islands in BCFT}},
  \href{https://arxiv.org/abs/2107.09083}{{\ttfamily 2107.09083}}.

\bibitem{Pedraza:2021cvx}
J.~F. Pedraza, A.~Svesko, W.~Sybesma and M.~R. Visser, \emph{{Semi-classical
  thermodynamics of quantum extremal surfaces in Jackiw-Teitelboim gravity}},
  \href{https://arxiv.org/abs/2107.10358}{{\ttfamily 2107.10358}}.

\bibitem{Iizuka:2021tut}
N.~Iizuka, A.~Miyata and T.~Ugajin, \emph{{A comment on a fine-grained
  description of evaporating black holes with baby universes}},
  \href{https://arxiv.org/abs/2111.07107}{{\ttfamily 2111.07107}}.

\bibitem{Miyata:2021qsm}
A.~Miyata and T.~Ugajin, \emph{{Entanglement between two evaporating black
  holes}},  \href{https://arxiv.org/abs/2111.11688}{{\ttfamily 2111.11688}}.

\bibitem{Gaberdiel:2021kkp}
M.~R. Gaberdiel, B.~Knighton and J.~Vo\v{s}mera, \emph{{D-branes in
  $\mathrm{AdS}_3\times \mathrm{S}^3\times \mathbb{T}^4$ at $k=1$ and their
  holographic duals}},  \href{https://arxiv.org/abs/2110.05509}{{\ttfamily
  2110.05509}}.

\bibitem{Uhlemann:2021itz}
C.~F. Uhlemann, \emph{{Information transfer with a twist}},
  \href{https://arxiv.org/abs/2111.11443}{{\ttfamily 2111.11443}}.

\bibitem{Collier:2021ngi}
S.~Collier, D.~Mazac and Y.~Wang, \emph{{Bootstrapping Boundaries and Branes}},
   \href{https://arxiv.org/abs/2112.00750}{{\ttfamily 2112.00750}}.

\bibitem{Hollowood:2021lsw}
T.~J. Hollowood, S.~P. Kumar, A.~Legramandi and N.~Talwar, \emph{{Grey-body
  Factors, Irreversibility and Multiple Island Saddles}},
  \href{https://arxiv.org/abs/2111.02248}{{\ttfamily 2111.02248}}.

\bibitem{emparan2021holographic}
R.~Emparan, A.~M. Frassino, M.~Sasieta and M.~Tomašević, \emph{Holographic
  complexity of quantum black holes},
  \href{https://arxiv.org/abs/2112.04860}{{\ttfamily 2112.04860}}.

\bibitem{omidi2021entropy}
F.~Omidi, \emph{Entropy of hawking radiation for two-sided hyperscaling
  violating black branes},  \href{https://arxiv.org/abs/2112.05890}{{\ttfamily
  2112.05890}}.

\bibitem{bhattacharya2021bath}
A.~Bhattacharya, A.~Bhattacharyya, P.~Nandy and A.~K. Patra, \emph{Bath
  deformations, islands and holographic complexity},
  \href{https://arxiv.org/abs/2112.06967}{{\ttfamily 2112.06967}}.

\bibitem{Merna}
A.~Karch, C.~Perez-Pardavila, M.~Riojas and M.~Youssef{\emph{, in preparation}
  (2023) }.

\bibitem{Almheiri:2020cfm}
A.~Almheiri, T.~Hartman, J.~Maldacena, E.~Shaghoulian and A.~Tajdini,
  \emph{{The entropy of Hawking radiation}},
  \href{https://arxiv.org/abs/2006.06872}{{\ttfamily 2006.06872}}.

\bibitem{Raju:2020smc}
S.~Raju, \emph{{Lessons from the Information Paradox}},
  \href{https://arxiv.org/abs/2012.05770}{{\ttfamily 2012.05770}}.

\bibitem{Raju:2021lwh}
S.~Raju, \emph{{Failure of the split property in gravity and the information
  paradox}},  \href{https://arxiv.org/abs/2110.05470}{{\ttfamily 2110.05470}}.

\bibitem{Liu:2020rrn}
H.~Liu and J.~Sonner, \emph{{Quantum many-body physics from a gravitational
  lens}}, \href{https://doi.org/10.1038/s42254-020-0225-1}{\emph{Nature Rev.
  Phys.} {\bfseries 2} (2020) 615}
  [\href{https://arxiv.org/abs/2004.06159}{{\ttfamily 2004.06159}}].

\bibitem{Nomura:2020ewg}
Y.~Nomura, \emph{{From the Black Hole Conundrum to the Structure of Quantum
  Gravity}}, \href{https://doi.org/10.1142/S021773232130007X}{\emph{Mod. Phys.
  Lett. A} {\bfseries 36} (2021) 2130007}
  [\href{https://arxiv.org/abs/2011.08707}{{\ttfamily 2011.08707}}].

\bibitem{Kibe:2021gtw}
T.~Kibe, P.~Mandayam and A.~Mukhopadhyay, \emph{{Holographic spacetime, black
  holes and quantum error correcting codes: A review}},
  \href{https://arxiv.org/abs/2110.14669}{{\ttfamily 2110.14669}}.

\bibitem{Geng_2022bhba}
H.~Geng, L.~Randall and E.~Swanson, \emph{{BCFT} in a black hole background: an
  analytical holographic model},
  \href{https://doi.org/10.1007/jhep12(2022)056}{\emph{Journal of High Energy
  Physics} {\bfseries 2022} (2022) }.

\bibitem{Geng:2021wcq}
H.~Geng, Y.~Nomura and H.-Y. Sun, \emph{{Information paradox and its resolution
  in de Sitter holography}},
  \href{https://doi.org/10.1103/PhysRevD.103.126004}{\emph{Phys. Rev. D}
  {\bfseries 103} (2021) 126004}
  [\href{https://arxiv.org/abs/2103.07477}{{\ttfamily 2103.07477}}].

\bibitem{Chen:2020uac}
H.~Z. Chen, R.~C. Myers, D.~Neuenfeld, I.~A. Reyes and J.~Sandor,
  \emph{{Quantum Extremal Islands Made Easy, Part I: Entanglement on the
  Brane}}, \href{https://doi.org/10.1007/JHEP10(2020)166}{\emph{JHEP}
  {\bfseries 10} (2020) 166}
  [\href{https://arxiv.org/abs/2006.04851}{{\ttfamily 2006.04851}}].

\bibitem{Chen:2020hmv}
H.~Z. Chen, R.~C. Myers, D.~Neuenfeld, I.~A. Reyes and J.~Sandor,
  \emph{{Quantum Extremal Islands Made Easy, Part II: Black Holes on the
  Brane}},  \href{https://arxiv.org/abs/2010.00018}{{\ttfamily 2010.00018}}.

\bibitem{asRong-Xin_Miao1}
R.-X. Miao, \emph{{Entanglement Island and Page Curve in Wedge Holography}},
  \href{https://arxiv.org/abs/2301.06285}{{\ttfamily 2301.06285}}.

\bibitem{Rong-Xin_Miao2}
R.-X. Miao, \emph{{Massless Entanglement Island in Wedge Holography}},
  \href{https://arxiv.org/abs/2212.07645}{{\ttfamily 2212.07645}}.

\bibitem{Liu_2022}
Y.~Liu, Z.-Y. Xian, C.~Peng and Y.~Ling, \emph{Black holes entangled by
  radiation}, \href{https://doi.org/10.1007/jhep09(2022)179}{\emph{Journal of
  High Energy Physics} {\bfseries 2022} (2022) }.

\bibitem{GengDGP}
H.~Geng, A.~Karch, C.~Perez-Pardavila, S.~Raju, L.~Randall, M.~Riojas,
  S.~Shashi et~al.{\emph{, in preparation} (2023) }.

\bibitem{Geng_2022jt}
H.~Geng, A.~Karch, C.~Perez-Pardavila, S.~Raju, L.~Randall, M.~Riojas and
  S.~Shashi, \emph{Jackiw-teitelboim gravity from the karch-randall
  braneworld},
  \href{https://doi.org/10.1103/physrevlett.129.231601}{\emph{Physical Review
  Letters} {\bfseries 129} (2022) }.

\bibitem{Geng:2021hlu}
H.~Geng, A.~Karch, C.~Perez-Pardavila, S.~Raju, L.~Randall, M.~Riojas and
  S.~Shashi, \emph{{Inconsistency of Islands in Theories with Long-Range
  Gravity}},  \href{https://arxiv.org/abs/2107.03390}{{\ttfamily 2107.03390}}.

\bibitem{McAvity:1995zd}
D.~McAvity and H.~Osborn, \emph{{Conformal field theories near a boundary in
  general dimensions}},
  \href{https://doi.org/10.1016/0550-3213(95)00476-9}{\emph{Nucl. Phys. B}
  {\bfseries 455} (1995) 522}
  [\href{https://arxiv.org/abs/cond-mat/9505127}{{\ttfamily
  cond-mat/9505127}}].

\bibitem{Porrati:2001gx}
M.~Porrati, \emph{{Mass and gauge invariance 4. Holography for the
  Karch-Randall model}},
  \href{https://doi.org/10.1103/PhysRevD.65.044015}{\emph{Phys. Rev. D}
  {\bfseries 65} (2002) 044015}
  [\href{https://arxiv.org/abs/hep-th/0109017}{{\ttfamily hep-th/0109017}}].

\bibitem{Porrati:2002dt}
M.~Porrati and A.~Starinets, \emph{{On the graviton selfenergy in AdS(4)}},
  \href{https://doi.org/10.1016/S0370-2693(02)01490-9}{\emph{Phys. Lett. B}
  {\bfseries 532} (2002) 48}
  [\href{https://arxiv.org/abs/hep-th/0201261}{{\ttfamily hep-th/0201261}}].

\bibitem{Porrati:2003sa}
M.~Porrati, \emph{{Higgs phenomenon for the graviton in ADS space}},
  \href{https://doi.org/10.1142/S0217732303011745}{\emph{Mod. Phys. Lett. A}
  {\bfseries 18} (2003) 1793}
  [\href{https://arxiv.org/abs/hep-th/0306253}{{\ttfamily hep-th/0306253}}].

\bibitem{Aharony:2006hz}
O.~Aharony, A.~B. Clark and A.~Karch, \emph{{The CFT/AdS correspondence,
  massive gravitons and a connectivity index conjecture}},
  \href{https://doi.org/10.1103/PhysRevD.74.086006}{\emph{Phys. Rev. D}
  {\bfseries 74} (2006) 086006}
  [\href{https://arxiv.org/abs/hep-th/0608089}{{\ttfamily hep-th/0608089}}].

\bibitem{Miemiec:2000eq}
A.~Miemiec, \emph{{A Power law for the lowest eigenvalue in localized massive
  gravity}},
  \href{https://doi.org/10.1002/1521-3978(200107)49:7<747::AID-PROP747>3.0.CO;2-T}{\emph{Fortsch.
  Phys.} {\bfseries 49} (2001) 747}
  [\href{https://arxiv.org/abs/hep-th/0011160}{{\ttfamily hep-th/0011160}}].

\bibitem{Schwartz:2000ip}
M.~D. Schwartz, \emph{{The Emergence of localized gravity}},
  \href{https://doi.org/10.1016/S0370-2693(01)00152-6}{\emph{Phys. Lett. B}
  {\bfseries 502} (2001) 223}
  [\href{https://arxiv.org/abs/hep-th/0011177}{{\ttfamily hep-th/0011177}}].

\bibitem{Geng:2020qvw}
H.~Geng and A.~Karch, \emph{{Massive islands}},
  \href{https://doi.org/10.1007/JHEP09(2020)121}{\emph{JHEP} {\bfseries 09}
  (2020) 121} [\href{https://arxiv.org/abs/2006.02438}{{\ttfamily
  2006.02438}}].

\end{thebibliography}\endgroup
\end{document}